\documentclass[11pt]{article}
\usepackage{graphicx}
\usepackage{amsfonts}
\usepackage{tikz}
	\usetikzlibrary{positioning}
\usepackage{amsmath,amssymb}
\usepackage{xcolor}
\usepackage{subcaption}
\usepackage{comment}
\usepackage{pdfpages}
\usepackage{multirow}

\usepackage{calligra} % for script and cursive characters like script r
\DeclareMathAlphabet{\mathcalligra}{T1}{calligra}{m}{n}
\DeclareFontShape{T1}{calligra}{m}{n}{<->s*[2.2]callig15}{}

\usepackage[margin=.65in]{geometry}

\definecolor{indigo}{RGB}{0,0,120}

\newcommand{\tl}[1]{\tilde{#1}}

\def\mod{\text{mod~}}

\def\lcm{\text{LCM}}

\newcommand{\beq}{\begin{equation}}
\newcommand{\eeq}{\end{equation}}
\newcommand{\beqs}{\begin{eqnarray}}
\newcommand{\eeqs}{\end{eqnarray}}

\newcommand{\ov}[1]{\frac{1}{#1}}

\newcommand{\un}[1]{\underline{#1}}

\def\fl{\noindent}

\def\al{\alpha}		
\def\g{\gamma} 		
 
\def\del{\delta}	
\def\D{\Delta}		
\def\eps{\epsilon}

\def\sig{\sigma}

\def\Sig{\Sigma}		
\def\tht{\theta}	
		
\def\om{\omega}

\def\vf{\varphi}

\usepackage{titlesec}
\titleformat{\section}{\normalsize\bfseries}{\thesection}{1em}{}
\titleformat{\subsection}{\small\bfseries}{\thesubsection}{1em}{}
\titleformat{\subsubsection}{\small\bfseries}{\thesubsubsection}{1em}{}

\newcount\colveccount  % macro for column vec eg. \colvec{2}{a}{b}
\newcommand*\colvec[1]{\global\colveccount#1  \begin{pmatrix} \colvecnext} \def\colvecnext#1{#1 \global\advance\colveccount-1
        \ifnum\colveccount>0 \\ \expandafter\colvecnext
        \else \end{pmatrix} \fi}

\newenvironment{smmat}
  {\left(\begin{smallmatrix}}
  {\end{smallmatrix}\right)}

\usepackage[colorlinks=true, linkcolor=indigo, citecolor=blue, urlcolor=indigo, backref=page]{hyperref}

\title{\hfill{\tt \small \href{https://arxiv.org/abs/2606.25766}{arXiv:2606.25766 [nlin.CD]}}\\
\Large Symbol sequences from three-rotor coincidences and their word-complexity}

\author{{\sc Govind S. Krishnaswami$^{a}$ and Anirudh Rameshan$^{a,b}$}
\\ \small
$^{a}$Physics Department, Chennai Mathematical Institute,  SIPCOT IT Park, Siruseri 603103, India\\ \small
$^{b}$UM-DAE Centre for Excellence in Basic Sciences, University of Mumbai, Vidyanagari, Mumbai 400098, India \\
\small
Email: {\tt govind@cmi.ac.in, anirudhr@cmi.ac.in}}

\date{October 30, 2026} 

\begin{document}

\maketitle

%---------------------
\vspace{-8mm}
\centering
\small Published in \href{https://doi.org/10.1063/5.0354972}{Chaos \textbf{36}, 103140 (2026)}
%---------------------

\abstract{\small In the three-rotor problem, three equally massive point particles move on a circle interacting via attractive pairwise cosine potentials. Rotors can represent superconducting phases of metallic segments in a chain of coupled Josephson junctions. We propose a digitization of the classical dynamics that records successive pair and triple coincidences of rotors using four symbols. Rotor coincidences correspond to boundaries in a disjoint partition of the configuration torus into cells where the rotors are ordered clockwise and anticlockwise. It is shown that isolated rotor coincidences must be crossings. Despite being a rather coarse digitization, we find that replacing trajectories by coincidence symbol sequences captures significant qualitative features of the dynamics through word statistics. Word-complexity $C_n$ measures the diversity of $n$-letter words in the symbol sequence while topological entropy governs asymptotic exponential growth of $C_n$. Sequences from periodic orbits have a word-complexity that saturates at the period. Ultra-high-energy trajectories with irrational `slope' are quasiperiodic. We show that they have zero entropy and $C_n = n+3$ by examining limiting slopes and by a mapping to Sturmian sequences. We examine their grammar rules and propose how their right-special words may be identified. On the other hand, numerical investigation of sequences from globally chaotic orbits leads us to conjecture an exponentially growing word-complexity $C_n = 3 \times 2^{n-1}$ with a topological entropy $\log 2$. We identify grammar rules that imply this conjectured word-complexity and model the resulting sequences as a subshift of finite type, unlike those at ultra-high energies which are not topological Markov chains.}

% Abstract: 240/250 words and 1702 characters

\scriptsize
\tableofcontents
\normalsize

% Keywords: three-rotors, rotor coincidence, symbol sequence, symbolic dynamics, word-complexity, Sturmian sequence, topological entropy, topological Markov shift, global chaos, quasiperiodic orbits, three body problems

\section*{}

{\bf \small A well-chosen discrete-time approximation to a continuous-time system can encode interesting properties of the dynamics while being easier to analyze. In symbolic dynamics, trajectories are replaced by sequences of symbols, each representing distinct physical possibilities. Symbols can label a partition of the configuration or state space. One of the first successful applications of symbolic dynamics was to geodesic flow on compact Riemann surfaces of constant negative curvature. More recently, the planar 3-body problem has been digitized by recording the identity of the middle body in successive collinearities. In this paper, we initiate a symbolic approximation to the 3-rotor problem where point particles move on a circle subject to attractive cosine interactions. The system shows order-chaos-order behavior with increasing energy and can model chains of coupled Josephson junctions. Symbol sequences are constructed by recording symbols from a four-letter alphabet labeling pair (1-2, 2-3, 3-1) and triple (1-2-3) rotor coincidences. The statistics of words that appear in these sequences enable us to distinguish between types of trajectories. In particular, the word-complexity $\boldsymbol{C_n}$ (number of distinct $\boldsymbol{n}$-letter words) and its exponential growth rate (topological entropy) allow us to quantify their chaotic nature.}

{\bf \small We examine symbol sequences from classes of periodic and ultra-high-energy (UHE) quasiperiodic orbits as well as chaotic ones from the energy band of global chaos. While the word-complexity of sequences of periodic orbits is bounded, that of UHE quasiperiodic and globally chaotic ones displays linear and (conjecturally) exponential growth. In fact, we conjecture that among our 3-rotor coincidence sequences, the latter are the aperiodic sequences with minimal and maximal topological entropy. The grammar of the aforementioned UHE sequences cannot be described by a finite number of forbidden words, implying that they do not form a subshift of finite type or a topological Markov chain. Nevertheless, we relate them to Sturmian sequences, which may be obtained by recording a pair of distinct symbols when a straight line with irrational slope cuts the horizontal and vertical lines in a square grid. We trace their linear word-complexity ($\boldsymbol{C_n = n + 3}$) to the uniqueness of the $\boldsymbol{n}$-letter word that admits two distinct suffixes and also develop an algorithm to identify such `right-special' words. By contrast, investigations into coincidence symbol sequences of chaotic orbits from the band of global chaos lead us to conjecture that their word-complexity is $\boldsymbol{C_n = 3 \times 2^{n-1}}$. We explain how this could arise from a pair of simple grammar rules that forbid triple coincidences and repetition of symbols. This allows us to model such chaotic coincidence sequences via a subshift of finite type and assign to them a topological entropy of $\boldsymbol{\log 2}$.}

%-----------------
\section{Introduction}
\label{s:intro}
%-----------------

An appropriately chosen discrete-time dynamics viewed as an approximation to a continuous-time system can capture key features of the latter while being more tractable. One can imagine several ways to set up a discrete-time dynamics: finite-difference approximation of differential equations, equal-time sampling of trajectories, Poincar\'e return map on a section, recording physically significant events along trajectories, etc., each with its advantages. For instance, E N Lorenz used iterations of a one-dimensional map \cite{lorenz-1963} to argue that the attractor in his system of ordinary differential equations modeling atmospheric convection cannot have any stable limit cycles. If the state space of the map is a finite alphabet, then we get a symbolic dynamical system. In pioneering work, J Hadamard \cite{hadamard-1898} associated a symbolic dynamics to geodesic flow on constant negative curvature compact Riemann surfaces. Other early contributors include G D Birkhoff \cite{birkhoff-book-1927}, M Morse and G A Hedlund \cite{morse-hedlund-1938,morse-hedlund-1940} and Smale \cite{smale-1965}. Symbolic dynamics and its application to physical systems continues to be an active area of research \cite{hao-1991,arioli-2002,hirata-mees-2003,matsuoka-hiraide-2015,day-2018,ampilova-soloviev-2019}. Some recent reviews may be found in \cite{lind-marcus-book-1995,fogg-book-2002,marcus-williams-2008,hao-zheng-book-2018,bruin-book-2022,hirata-amigo-2023}. 

In the three-body problem, recording the identity of the middle body in successive syzygies (collinearities) has been a fruitful approach in attempts to develop a complete symbolic dynamics for the planar and Euler versions of the problem \cite{montgomery-2002,dullin-montgomery-2016}. Even if a symbolic digitization is not  topologically conjugate to the continuous dynamical system, the presence of fewer degrees of freedom can allow the former to serve as a simpler context to approximately understand some statistical or other features of the original dynamics. For instance, it is computationally less intensive to evaluate the word-complexity or topological entropy \cite{adler-top-entropy-2008} in a symbolic digitization than to find Lyapunov exponents or the Kolmogorov--Sinai entropy of the smooth flow. Quite apart from its use as an approximation, a good digitization can lead to an interesting dynamical system in its own right.

%Nevertheless, there can be relations between these quantities \cite{catalan-2019}.

In this paper, we initiate an investigation of a symbolic dynamical approximation to the classical three-rotor problem \cite{gsk-hs-2019} with rotor coincidences playing the role that syzygies played in the three-body problem. By replacing trajectories with symbol sequences from an alphabet labeling pair and triple coincidences of rotors, we propose a digitization of the dynamics that (i) uses a small number of symbols, (ii) is intrinsic to the dynamics (does not involve a choice of time step), (iii) is based on interesting physical events, (iv) appears largely capable of distinguishing periodic, quasiperiodic and chaotic orbits despite being a rather coarse discretization (compared, say, to equal-time sampling) and (v) allows us to quantify their complexity.

\vspace{4pt}

\fl {\bf Setup of the 3-rotor problem.} In the 3-rotor problem, three point particles of mass $m$ move on a circle of radius $r$ subject to attractive cosine inter-particle potentials of strength $g$. If $\tht_1, \tht_2, \tht_3$ are the rotor angles, then the Hamiltonian is
	\beq
	H_{\rm tot} = \sum_{i=1}^{3} \left\{\dfrac{\pi_i^2}{2mr^2} + g [1 - \cos (\tht_i - \tht_{i+1})]\right\},
	\label{e:hamiltonian-full}
	\eeq
where $\tht_4 \equiv \tht_1$ and $\pi_i = mr^2 \dot \tht_i$ are the conjugate angular momenta. The rotor angles can model superconducting phases in distinct metallic segments of a chain of coupled Josephson junctions \cite{orlando-et-al-1999,gsk-ay-2023}. Thus, the rotors are treated as identical but distinguishable particles that can pass through each other. It is convenient to define center of mass (CM) and relative (Josephson) angles
	\beq
	\vf_0 = (\tht_1 + \tht_2 + \tht_3)/3, \quad 
	\vf_1 = \tht_1 - \tht_2 \quad 
	\text{and} \quad
	\vf_2 = \tht_2 - \tht_3,
	\label{e:cm-rel-angles}
	\eeq
in terms of which the kinetic and potential energies are
	\beqs
	T &=& \dfrac{3}{2} mr^2 \dot \vf_0^2  + \dfrac{1}{3} mr^2 \left[ \dot \vf_1^2 + \dot \vf_2^2 + \dot \vf_1 \dot \vf_2\right] \quad \text{and} \cr
	V &=& g[3- \cos \vf_1 - \cos \vf_2 - \cos(\vf_1 +\vf_2)].
	\eeqs
While $3mr^2 \ddot \vf_0 = 0$, the relative angles evolve independently of the CM angle:
	\beq
	mr^2 \ddot \vf_1 = - g [ 2 \sin \vf_1 - \sin \vf_2 + \sin (\vf_1 + \vf_2)] \quad 
	\text{and} \quad 
	1 \leftrightarrow 2.
	\label{e:phi-eom}
	\eeq
They may be taken to be $2\pi$ periodic. The Hamiltonian becomes 
	\beq
	H_{\rm tot} = \dfrac{p_0^2}{6mr^2} + E \quad \text{where} \quad E = \dfrac{p_1^2 + p_2^2 - p_1 p_2}{mr^2} + V (\vf_1, \vf_2)
	\eeq
is the relative energy and the conjugate angular momenta are
	\beq
	p_0 = 3mr^2\dot\vf_0, \quad 
	p_1 = \dfrac{mr^2}{3} (2\dot\vf_1 + \dot\vf_2) \quad 
	\text{and} \quad 
	p_2 = \dfrac{mr^2}{3} (2\dot\vf_2 + \dot\vf_1).
	\label{e:mom-conj-cm-rel-ang}
	\eeq
%The corresponding Hamilton's equations are $\dot \vf_0 = p_0/3mr^3$, $\dot p_0 = 0$, 
%	\beq
%	\dot \vf_1 = (2p_1-p_2)/mr^2, \quad 
%	\dot p_1 = -g[\sin \vf_1 + \sin(\vf_1 + \vf_2)] \quad
%	\text{and} \quad
%	1 \leftrightarrow 2.
%	\label{e:hamilton-eom}
%	\eeq
The relative motion on the $\vf_1$-$\vf_2$ configuration 2-torus admits only one conserved quantity, with $E/g$ serving as a dimensionless control parameter. The system is integrable in both the low and high energy limits, going from small oscillations around the ground state of three coincident rotors to uniform rotation of rotors. In Refs.~\cite{gsk-hs-2019,hs-2020}, Euler and Lagrange-type families of periodic solutions (pendula and isosceles breathers) and periodic choreographies were found. Pendula alternate between stable and unstable phases as the energy is varied, with stability transitions accumulating from either side at $E = 4 g$. These stability transitions have been shown \cite{gsk-ay-2023} to be associated with isochronous and period doubling bifurcations at which new families of periodic trajectories are born. What is more, the system shows order-chaos-order behavior as $E$ increases from zero to infinity. The crossover to widespread chaos occurs around $E = 4g$ and is associated with spontaneous breaking of discrete symmetries on Poincar\'e sections. Additionally, there is a band of seemingly global chaos for $5.33 g \lesssim E \lesssim 5.6 g$ where ergodic behavior and mixing have been reported \cite{gsk-hs-2020}. Quantum manifestations of regularity and chaos have also been studied \cite{gsk-hs-2024}.

%---------------
\subsection{Organization and summary of results}
\label{s:summary-results}
%---------------

We begin in \S \ref{s:rot-ord-rot-coincidence-sym-seq} by assigning symbol sequences to three-rotor trajectories using the distinguishable nature of rotors 1, 2 and 3 with angular coordinates $\tht_1$, $\tht_2$ and $\tht_3$. By recording the symbols 1, 2, 3 and 0 at each successive pair (2-3, 3-1, 1-2) and triple (1-2-3) coincidence of rotors, we obtain the symbol sequence for an orbit on the configuration space for relative motion. Interestingly, we show that the clockwise/anticlockwise order of rotors around the circle must change at an isolated meeting of rotors. Rotor-coincidence sequences turn out to capture many qualitative properties of periodic, quasiperiodic and chaotic trajectories with a modest-sized (four-letter) alphabet. These coincidence sequences furnish a physically well-motivated but rather coarse digitization of trajectories compared to equal-time sampling or Poincar\'e sections. Despite containing less information, they have the nice feature of not depending on an arbitrary time step or choice of Poincar\'e surface. In \S \ref{s:order-coinc-conditions}, we derive convenient formulas to determine rotor order and coincidences given the relative angles $\vf_1$ and $\vf_2$ between rotors. Rotor orders and coincidences together furnish a disjoint partition of the toroidal $\vf_1$-$\vf_2$ configuration space into a pair of cells along with their boundaries. 

In \S \ref{s:seq-space-shift-map-word-stat}, we introduce the coincidence sequence spaces and shift maps which are the discrete analogs of families of 3-rotor trajectories and time evolution along them. Useful statistical quantifiers of the nature of symbol sequences are defined. Word-complexity $C_n$ is the number of distinct $n$-letter words ($n$-words) in a symbol sequence. It quantifies the complexity of the sequence representing a trajectory, while the word frequency is a more refined measure. Topological entropy governs the asymptotic growth of word-complexity as $n \to \infty$. Section \ref{s:symb-seq-static-periodic} contains examples of symbol sequences for static solutions and known classes of periodic trajectories: pendula (where a pair of rotors are stuck together), isosceles breathers, small oscillations around the ground state and choreographies. 
Their word-complexities are obtained in \S \ref{s:static-periodic-word-complexity}. In \S \ref{s:pair-coinc-cross}, we show that rotors must cross at an isolated meeting of two rotors while in \S \ref{s:triple-coinc-cross} we argue that the order of rotors must change at an isolated triple coincidence. 

In \S \ref{s:UHE-ss}, we consider an ultra-high-energy (UHE) limit where the kinetic energy dominates interactions and trajectories are straight lines on the $\vf_1$-$\vf_2$ torus. They are periodic/quasiperiodic according as their slope $\sig = \dot \vf_2/\dot \vf_1$ is rational/irrational. Various structural features of such trajectories vis-\`a-vis our partitioning of the $\vf_1$-$\vf_2$ torus are identified in \S \ref{s:structure-traj-sym-seq-UHE}, along with a formula for the period of the symbol sequence when $\sig$ is rational. The rotor permutation symmetry is shown to allow us to restrict attention to $\sig > 1$. In \S \ref{s:uhe-periodic}, we examine the word statistics of UHE period-$p$ rotor-coincidence symbol sequences. Their word-complexity $C_n$ initially grows (in a piecewise linear manner) but once the word length $n \geq p$, it saturates at $C_n = p$ leading to vanishing topological entropy. Symbol sequences from UHE quasiperiodic orbits also have zero entropy, but are much richer and are examined in \S \ref{s:uhe-quasiperiodic}. Based on examples, in \S \ref{s:conjectures-uhe-quasiperiodic}, we propose that UHE quasiperiodic trajectories have a linear word-complexity $C_n = n+3$. This would be explained if there is precisely one $n$-word in such a symbol sequence that can be suffixed by a letter in exactly two ways to get an $(n+1)$-word. We also show how this unique `right-special' $n$-word may be identified. In \S \ref{s:gram-rule-uhe}, we find a (partial) list of grammar rules governing UHE symbol sequences, which we argue cannot form a topological Markov chain (or subshift of finite type). In \S \ref{s:asymp-argue-Cn-formula}, the formula $C_n = n+3$ is established in the limiting regimes $\sig \to \infty$ and $\sig \to 1$. In \S \ref{s:sturmian}, we find a mapping of UHE quasiperiodic rotor-coincidence sequences to Sturmian sequences, which are aperiodic binary sequences with minimal word-complexity \cite{morse-hedlund-1940,lind-marcus-book-1995,bruin-book-2022,lothaire-2002}. We use properties of the latter to deduce our conjectured word-complexity formula $C_n = n+3$ for all $\sig$. Interestingly, Sturmian sequences have also found application to Euler's integrable three-body problem of two fixed centers \cite{dullin-montgomery-2016} and to the H\'enon map \cite{dullin-meiss-stirling-2005}. Although they have zero entropy, these sequences cannot be defined by a finite number of forbidden words, which is consistent with our observation that UHE symbol sequences cannot form a topological Markov chain. We end this section by briefly touching upon the word-frequency of UHE quasiperiodic sequences in \S \ref{s:word-freq-uhe-quasiper}.

In \S \ref{s:word-cx-global-chaotic-band}, we turn from ultra-high energies to the conjectured band of global chaos $5.33 g \lesssim E \lesssim 5.6g$ and examine the word-complexity of symbol sequences of numerically obtained chaotic trajectories of increasing length (up to $t \sim 6 \times 10^6$ in units where $m = r = g = 1$ or sequence length $l = 6.72 \times 10^6$). The apparent exponential growth of $C_n$ with $n$ makes truncation effects significant even for moderate $n$ while the slow approach of $C_n$ to asymptotic values with increasing $l$ makes it challenging to extract asymptotic behavior. Nevertheless, extrapolating from available data, we conjecture that $C_n = 3 \times 2^{n-1}$ for rotor-coincidence symbol sequences of chaotic trajectories from the band of global chaos. This corresponds to a topological entropy $h = \log 2$. Examination of these sequences leads us to conjecture that they are defined by two rules: absence of 0s and symbol repetitions. We encode these rules via a 0-1 adjacency matrix that defines a topological Markov chain, unlike the coincidence sequences of quasiperiodic orbits at ultra-high energies. We conclude in \S \ref{s:discussion} with a brief discussion. The initial portion of this work was reported in the Masters thesis \cite{anirudh-rameshan-thesis}.

%-----------------
\section{Symbol sequences from 3-rotor dynamics}
\label{s:symb-seq}
%-----------------

%-----------------
\subsection{Rotor-order and rotor-coincidence symbol sequences}
\label{s:rot-ord-rot-coincidence-sym-seq}
%-----------------

\begin{figure}
	\centering
\includegraphics[width=0.35\textwidth]{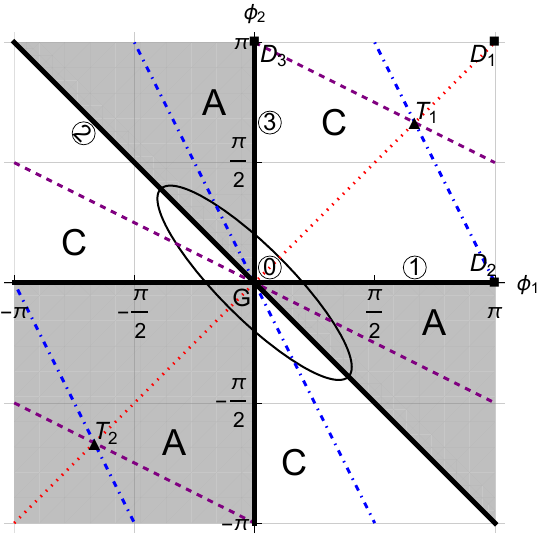}
	\caption{\small Fundamental domain of $\vf_1$-$\vf_2$ configuration torus. The rotors are ordered anticlockwise (A) and clockwise (C) in the shaded and unshaded regions. The thick black boundaries correspond to pairwise coincidences (1, 2 and 3), and triple coincidence (0) at the origin. The static solutions G, D and T are indicated. Librational and rotational periodic pendula lie along the coincidence lines 1, 2 and 3, while the corresponding isosceles breathers lie along dotted red, dot-dashed blue and dashed purple lines (near G, they have angular momentum $L_z = 0$). The ellipse, traversed clockwise ($L_z < 0$) or anticlockwise ($L_z > 0$), represents a magnified version of two low-energy  nonrotating choreographies.}
	\label{f:A-C-regions-phi1-phi2-square}
\end{figure}

\textbf{Symbol sequence based on rotor order.} A simple way of assigning a symbol sequence to a trajectory is to partition the $\vf_1$-$\vf_2$ configuration torus into two cells labeled C and A depending on the order of the rotors. Rotors are clockwise (C) if their order is 1-2-3 clockwise around the circle and anticlockwise (A) if the order is 1-2-3 anticlockwise. The boundaries of C and A correspond to configurations with two or three coincident rotors. The cells C and A along with their boundaries form the disjoint partition of the torus shown in Fig.~\ref{f:A-C-regions-phi1-phi2-square}. Given a trajectory $\g$ on this torus, we may associate a symbol sequence to it in the CA `rotor-order' alphabet by recording the successive cells through which the trajectory passes. If the trajectory is confined to the boundary between cells, the symbol sequence is empty. We will show in \S \ref{s:coinc-and-cross} that any isolated coincidence leads to a change in rotor order. Consequently, we cannot have a repetition of symbols (CC or AA). Hence, a symbol sequence must be alternating: empty, terminating (e.g., C, A) or (semi-)infinite (e.g., $\cdots$ACAC$\cdots$). However, these symbol sequences cannot effectively distinguish between periodic, quasiperiodic and chaotic trajectories.

A way to circumvent this shortcoming is to obtain symbol sequences in the CA alphabet by sampling trajectories at (say) equal intervals of time, thereby allowing for repetition of letters. The resulting symbol sequences (e.g., $\cdots$ACAAACCA$\cdots$) would have information on the time spent in each cell in addition to the order of cells traversed. However, this would involve a somewhat artificial choice of time step. In what follows, we propose a more nuanced way of recording changes in rotor order that furnishes a digitization with a small number of symbols but nevertheless captures qualitative properties of the dynamics.

\vspace{5pt}

\fl \textbf{Symbol sequence based on coincidences.} Consider a trajectory $\g$ on the $\vf_1$-$\vf_2$ configuration space starting somewhere at $t = 0$. We will maintain a symbolic record of rotor coincidences while ignoring the time elapsed between successive coincidences. In particular, our sequences will be defined in terms of configuration space trajectories and we will not discuss a partition of the phase space.

We first deal with isolated coincidences where rotors coincide at one instant $t_c$ and are separate both immediately before and after $t_c$. We associate the symbol 1 to a pair coincidence of rotors 2 and 3. Similarly, the symbols 2 and 3 are associated with 3-1 and 1-2 pair coincidences. In pair coincidences, it is assumed that the third rotor does not coincide with the participating pair. A triple coincidence is assigned the symbol 0. Thus, our `rotor-coincidence' alphabet ${\cal A} = \{ 0, 1, 2, 3 \}$ has four letters. The rotor-coincidence symbol sequence of trajectory $\g$ is the string of letters $s(\g)$ associated with each successive coincidence. While pair coincidences are routine, triple coincidences require fine-tuned initial conditions, so the symbol 0 will not occur in generic trajectories. Since trajectories can be extended indefinitely backward and forward in time, we may associate a bi-infinite symbol sequence to an extended trajectory $\g$. For instance, we may have $s(\g) = \cdots 132.132 \cdots$. The full stop separates coincidences before $t = 0$ from those that happen when $t \geq 0$. If the initial time is not relevant, we omit the punctuation mark. When coincidences are not isolated, i.e., if two or three rotors coincide for some duration, we record the symbol corresponding to the coincidence only once during this period. For instance, suppose rotors 1 and 2 are coincident for $0 \leq t \leq t_1$ and are met for the first time by rotor 3 at $t_1$, then $s(\g) = .30 \cdots$. Interestingly, we will see in \S \ref{s:pair-coinc-cross} and \S \ref{s:triple-coinc-cross} that if two or three rotors are together for some duration, then they must be coincident at all times in the past and future.

\vspace{5pt}

\fl \textbf{Switching alphabets and rotor exchange symmetry.} (a) It will be shown in \S \ref{s:coinc-and-cross} that the CA symbol sequence for a trajectory can be deduced from that in the 0123 alphabet when the initial rotor order is known. However, the converse is not true. (b) Due to the rotor exchange symmetry \cite{gsk-hs-2024} of the Hamiltonian (\ref{e:hamiltonian-full}), the effect of rotor exchange is easily dealt with. Suppose trajectory $\g'$ is obtained by exchanging rotors $i$ and $j$ in the initial state of $\g$. Then $s(\g')$ is obtained from $s(\g)$ by exchanging symbols $i$ and $j$ in $s(\g)$. On the other hand, in the CA alphabet, any exchange of rotors leads to an interchange of symbols C and A (see Table~\ref{t:rotor-exchanges}).

\vspace{5pt}

{\fl \bf Extension to four or more rotors.} Although not pursued here, rotor-order and rotor-coincidence sequences may be extended to $n$ rotors. For four rotors, we would record pair, triple and quadruple coincidences using an alphabet of $6+4+1 = 11$ symbols, although only the six pair coincidence symbols will appear in typical trajectories. For four rotors, there are six distinct rotor orders, which can be taken as $1234, 1243, 1324, 1342, 1423, 1432$ clockwise. In the $n$-rotor problem, there are (a) $2^n - n - 1$ types of rotor coincidences, of which only the $n(n-1)/2$ pair coincidences are generic and (b) $(n-1)!$ distinct orderings of rotors around the circle, obtainable by fixing one rotor (say, rotor 1) and permuting the rest.

% In the formula for the above ellipse, we took $E_1 = E_2= 2g$.

%-----------------
\subsection{Conditions for rotor orders and coincidences}
\label{s:order-coinc-conditions}
%-----------------

Here, we express the conditions for clockwise (C) and anticlockwise (A) rotor orders and for coincidences in terms of the relative angles $\vf_1 = \tht_1 - \tht_2$ and $\vf_2 = \tht_2 - \tht_3$. The partitioning of the configuration torus into corresponding C and A cells and their boundaries is also obtained. 

\begin{figure}[ht]
	\centering
	\includegraphics[width=0.30\textwidth]{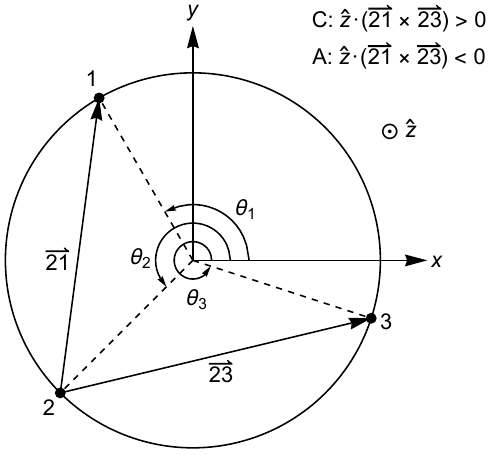}
	\caption{\small Rotor positions and inter-rotor vectors to find the rotor order (anticlockwise in this example).}
	\label{f:3rotor-with-vec}
\end{figure}

The order of rotors is determined by the sign of a certain scalar triple product. Suppose the rotors lie on the $x$-$y$ plane with $\hat z$ pointing out of the plane as in Fig.~\ref{f:3rotor-with-vec}. Let $\overrightarrow{21}$ be the vector from rotor 2 to rotor 1 and so forth. Then, the rotor order is
\beq
	\text{C or A according as} \quad \hat z \cdot (\overrightarrow{21} \times \overrightarrow{23} ) \gtrless 0.
	\eeq
Coincidences occur when the triple product vanishes. These conditions may be expressed purely in terms of the relative angles. In fact, using
	\beqs
	\overrightarrow{21} &=& r(\cos \tht_1-\cos \tht_2) \hat x + r(\sin \tht_1 - \sin \tht_2) \hat y \quad
	\text{and} \cr
	\overrightarrow{23} &=& r(\cos \tht_3-\cos \tht_2) \hat x + r(\sin \tht_3 - \sin \tht_2) \hat y,
	\eeqs
where $r$ is the radius of the circle, we arrive at the conditions
	\beq
	\text{C or A according as} \quad \sin \varphi_1 + \sin \varphi_2 - \sin (\varphi_1 + \varphi_2) \gtrless 0.
	\eeq
The C and A regions of the $\vf_1$-$\vf_2$ torus are indicated in Fig.~\ref{f:A-C-regions-phi1-phi2-square}. Coincidences occur along the boundary between C and A, i.e., when the inequalities are saturated. They are of four sorts:

\begin{itemize}

\item[(0)] Triple coincidence when $\vf_1 \equiv \vf_2 \equiv 0$ $\mod 2\pi$ or equivalently $\sin(\vf_1/2) = \sin (\vf_2/2) = 0$. 

\item[(1)] Purely 2-3 coincidence when $\vf_2 \equiv 0$ $\mod 2\pi$ but $\vf_1 \not\equiv 0$ $\mod 2\pi$ or equivalently $\sin(\vf_2/2)$ = 0 but $\sin (\vf_1/2) \ne 0$. 

\item[(2)] Purely 3-1 coincidence when $\vf_1 + \vf_2 \equiv 0$ $\mod 2\pi$ but $\vf_1 \not\equiv 0$ $\mod 2\pi$ or equivalently $\sin((\vf_1 + \vf_2)/2) = 0$ but $\sin(\vf_1/2) \ne 0$. 

\item[(3)] Purely 1-2 coincidence when $\vf_1 \equiv 0$ $\mod 2\pi$ but $\vf_2 \not\equiv 0$ $\mod 2\pi$ or equivalently $\sin(\vf_1/2)$ = 0 but $\sin(\vf_2/2) \ne 0$.
\end{itemize}
Notice that these conditions are expressed purely in terms of relative angles and do not involve the center of mass angle $\vf_0 = (\tht_1 + \tht_2 + \tht_3)/3$. The formulation in terms of the sine is helpful in detecting coincidences numerically. In practice, we replace inequalities such as $\sin(\vf_2/2) \ne 0$ that appear in the condition for pair coincidences with $|\sin(\vf_2/2)| > \del$, where $\del = 10^{-8}$ is a tolerance. Similarly, for triple coincidences, we replace one of the two conditions by the inequality $|\sin(\vf_1/2)| < \del$.

%--------
\subsection{Sequence spaces, shift map and word statistics}
\label{s:seq-space-shift-map-word-stat}
%--------

We now introduce some useful concepts that we employ in our study of symbol sequences arising from 3-rotor dynamics. These include (i) sequence spaces and shift maps, which provide the arena for symbolic dynamics and (ii) word-complexity, word-frequency and topological entropy, which allow us to quantify the complexity of trajectories.

\vspace{5pt}

\fl \textbf{Coincidence sequence spaces, subshifts and SFT.} If symbols are recorded only at times beginning with an initial time (say, $t = 0$) then we get semi-infinite sequences rather than bi-infinite ones. However, not every symbol sequence may arise from three-rotor dynamics, so the semi-infinite and bi-infinite sequences in the alphabet ${\cal A} = \{ 0, 1, 2, 3 \}$ are subsets $X_+$ and $X$ of the {\it full} sequence spaces $\Sigma = {\cal A}^\mathbb{N}$ and ${\cal A}^\mathbb{Z}$, where $\mathbb{N}$ and $\mathbb{Z}$ are the natural numbers and integers. {\it Symbolic dynamics} is given by the left-shift maps: $\tau(. s_1 s_2 s_3 \cdots) = .s_2 s_3 \cdots$ and $\tau(\cdots s_{-2} s_{-1} . s_0 s_1 s_2 \cdots) = \cdots s_{-1} s_0 . s_1 s_2 \cdots$, where the full stop is used to demarcate the current time rather than $t = 0$. In other words, $\tau(s)_i = s_{i+1}$ for $i \in \mathbb{N}$ or $\mathbb{Z}$. For bi-infinite sequences, we also have backward time evolution which is given by the right-shift or $\tau^{-1}$. A full sequence space $\Sig$ along with the shift map $\tau$ is called the full-shift $(\Sig,\tau)$ \cite{bruin-book-2022}. Since the three-rotor system is time-translation invariant and any initial state can be evolved indefinitely both forward and backward in time, the sequence spaces $X_+$ and $X$ are both strongly shift-invariant: $\tau(X_+) = X_+$ and $\tau(X) = X$ (in fact, $\tau^{-1}(X) = X$ as well). A closed, shift-invariant subset of a full-shift is called a {\it subshift}. Thus, $(X_+,\tau)$ and $(X,\tau)$ are subshifts of the full-shifts. A subshift is said to be of {\it finite type} (an SFT) or a {\it topological Markov shift} if it consists of sequences that avoid a finite set of forbidden words.

\vspace{5pt}

\fl \textbf{Word-complexity and topological entropy.} The diversity of words that appear in a symbol sequence $s(\g)$ can be used to quantify the complexity of rotor crossings in $\g$. The {\it word-complexity} $C_n(s(\g))$ is the number of distinct $n$-letter words (or $n$-words) in $s(\g)$, while the {\it topological entropy} $h(s)$ measures its asymptotic growth:
	\beq
	h(s) = \lim\limits_{n \to \infty} n^{-1} \log C_n(s).
	\eeq
If the word-complexity grows as a power rather than exponentially, then a useful quantifier is the {\it power entropy} $h_{\rm pow}(s) = \lim_{n \to \infty} \log C_n(s) / \log n$ \cite{bruin-book-2022}. In \S \ref{s:static-periodic-word-complexity}, \S \ref{s:UHE-ss} and \S \ref{s:word-cx-global-chaotic-band} we will study the word-complexity of symbol sequences (in the 0123 alphabet) from some static, periodic, quasiperiodic and chaotic orbits. We note that the maximum possible number of $n$-words is $C_n^{\rm max} = 4^n$, corresponding to a maximum possible topological entropy $h_{\rm max} = \log 4 = 2 \log 2$.

\vspace{5pt}

\fl {\bf Word-frequency.} The word-frequency \cite{bruin-book-2022} $f_w(s)$ of a word $w$ in a sequence $s = s_0 s_1 s_2 \cdots$ is the frequency with which the word appears in $s$ and is defined as 
	\beq
	f_w(s) = \lim_{n \to \infty} \ov{n} \# \{ 0 \le i < n : s_i \cdots s_{i+|w|-1} = w \}
	\eeq
where $\# \{ \cdots \}$ is the cardinality and $|w|$ is the length of the word $w$. The word frequencies of some periodic, quasiperiodic and chaotic orbits are discussed in \S \ref{s:uhe-periodic}, \S \ref{s:word-freq-uhe-quasiper} and \S \ref{s:word-cx-global-chaotic-band}.

%-----------------
\section{Static and periodic orbits}
\label{s:static-periodic}
%-----------------

In this section, we identify the symbol sequences of all static and some classes of periodic orbits for the relative motion of three rotors. Their word-complexities saturate at the period of the corresponding sequence.

%-----------------
\subsection{Symbol sequences for static and simple periodic orbits}
\label{s:symb-seq-static-periodic}
%-----------------

Static solutions have symbol sequences of length zero or one. In \cite{gsk-hs-2019} several classes of periodic orbits of the 3-rotor problem were found. We show here that pendula, breathers and small oscillations around the ground state, including low-energy choreographies, have periodic symbol sequences with period up to three in the 0123 alphabet. These examples and the accompanying formulas will help us formulate a conjecture on the nature of rotor coincidences and show in \S \ref{s:coinc-and-cross} that isolated coincidences must involve a crossing of rotors.

\vspace{5pt}

\fl {\bf Static solutions.} Static solutions for the relative motion are of three sorts: G, D and T (see \S III A of \cite{gsk-hs-2019} and Fig.~\ref{f:A-C-regions-phi1-phi2-square}). In the ground state G, $\vf_1 \equiv \vf_2 \equiv 0$ so that all three rotors are coincident at all times. Consequently, G has the symbol sequence 0. In the diagonal states D, two rotors are always coincident while the third is diametrically opposite, so that $s(\text{D}) = 1$, 2 or 3, depending on which two rotors coincide. In the triangle states T, the rotors are always at the vertices of an equilateral triangle, resulting in an empty symbol sequence.

\vspace{5pt}

\fl {\bf Periodic pendula.} Here, two rotors form a molecule while the third oscillates about their center of mass like a pendulum (see \S IV A of \cite{gsk-hs-2019}). In Fig.~\ref{f:A-C-regions-phi1-phi2-square}, pendula lie along the coincidence lines $\vf_1 = 0$, $\vf_2 = 0$ and $\vf_1 + \vf_2 =0$ (modulo 2$\pi$). Nonstatic pendula undergo periodic triple coincidences, leading to the sequences given in Table~\ref{t:static-periodic-symb-seq}, depending on which pair form a molecule. The symbol sequence does not distinguish librational from rotational pendula.

\vspace{5pt}

\fl {\bf Isosceles breathers.} Here, one rotor remains at the center of mass while the other two oscillate (librate/rotate) symmetrically (see \S IV B of \cite{gsk-hs-2019}). In Fig.~\ref{f:A-C-regions-phi1-phi2-square}, breathers lie along the lines $\vf_1 = \vf_2$, $\vf_1 = -2 \vf_2$ and $\vf_2 = -2 \vf_1$ (modulo $2\pi$). The librational breathers (LG and LD) involve oscillations around G and D, and feature only triple and pairwise coincidences, respectively. In rotational (R) breathers, pairwise coincidences alternate with triple coincidences. The rotor-order is reversed at each of these coincidences. Their symbol sequences are given in Table~\ref{t:static-periodic-symb-seq}.

\begin{table}[ht]
	\centering
	\begin{tabular}{|l|c|c|c|}
		\hline
		\multicolumn{2}{|c|}{\multirow{2}{*}{\textbf{Solution type}}} & \multicolumn{2}{|c|}{\textbf{Symbol sequence in}} \\
		\cline{3-4}
		\multicolumn{1}{|c}{} &  & \textbf{CA alphabet} & \textbf{0123 alphabet} \\ 
		\hline
		\multirow{3}{*}{Static} & G & \multirow{2}{*}{empty} & 0 \\
		\cline{2-2} \cline{4-4}
		& D & & 1 or 2 or 3 \\
		\cline{2-4}
		& T & C or A & empty \\		
		\hline
		\multicolumn{2}{|l|}{Pendula} & empty & $\cdots 0101 \cdots$ or $\cdots 0202 \cdots$ or $\cdots 0303 \cdots$ \\
		\hline
		\multirow{3}{*}{Breathers} & LG & \multirow{3}{*}{$\cdots$ CACA $\cdots$} & $\cdots 0000 \cdots$ \\
		\cline{2-2} \cline{4-4}
		& LD & & $\cdots 1111 \cdots$ or $\cdots 2222 \cdots$ or $\cdots 3333 \cdots$ \\
		\cline{2-2} \cline{4-4}
		& R & & $\cdots 0101 \cdots$ or $\cdots 0202 \cdots$ or $\cdots 0303 \cdots$ \\
		\hline
		Small osc. around G & $L_z > 0$ & \multirow{3}{*}{$\cdots$ CACA $\cdots$} & $\cdots 321321 \cdots$ (anticlockwise)\\
		\cline{2-2} \cline{4-4}
		E.g., choreographies & $L_z < 0$ & & $\cdots 312312 \cdots$ (clockwise)\\
		\cline{2-2} \cline{4-4}
		E.g., breathers & \multirow{2}{*}{$L_z = 0$} & & $\cdots 0000 \cdots$ \\
		\cline{3-4}
		E.g., pendula & & empty &	$\cdots 0101 \cdots$ or $\cdots 0202 \cdots$ or $\cdots 0303 \cdots$ \\
		\hline	
	\end{tabular}
	\caption{\small Symbol sequences of static and some simple periodic solutions. Notice that pendula and rotational breathers can have the same sequence, say, $\cdots 0101 \cdots$. Our digitization does not distinguish between eternal and isolated coincidences: the 1s in pendula are eternal while those in breathers are isolated. However, this type of ambiguity is rare, since pendula are the only trajectories with eternal coincidences (see \S IV A of \cite{gsk-hs-2019}).}
	\label{t:static-periodic-symb-seq}
\end{table}

\fl {\bf Small oscillations around ground state G.} At low relative energies $E \ll g$, the equations of motion (\ref{e:phi-eom}) reduce to those of a pair of oscillators $\ddot \vf_{1,2} = - \om_0^2 \vf_{1,2}$ with the {\it same} angular frequency $\om_0 = \sqrt{3g/m r^2}$ (see \S III C of \cite{gsk-hs-2019}). If $E_{1,2} \ll g$ denote the mode energies, then
	\beq
	\vf_{1,2} = \sqrt{2 E_{1,2}/3g} \: \sin(\om_0 (t - t_{1,2}))
	\label{e:soln-sm-osc}
	\eeq
where $\om_0 t_{1,2}$ are initial phases. They correspond to elliptical trajectories, as shown in Fig~\ref{f:A-C-regions-phi1-phi2-square}. We will assume that $E_1$ and $E_2$ are not both zero, in which case we recover the static solution at G. It is convenient to introduce polar coordinates $\rho = \sqrt{\vf_1^2 + \vf_2^2}$ and $\psi = \arctan(\vf_2/\vf_1)$ with
	\beq
	\dot \psi = \frac{\vf_1 \dot \vf_2 - \vf_2 \dot \vf_1}{\vf_1^2 + \vf_2^2} = \frac{L_z}{mr^2 \rho^2}
	\quad \text{where} \quad
	L_z = mr^2 (\vf_1 \dot \vf_2 - \vf_2 \dot \vf_1)
	= \frac{2}{\om_0} \sqrt{E_1 E_2} \sin(\om_0 (t_2 - t_1))
	\label{e:ang-mom-sm-osc-and-psi-dot}
	\eeq
is the conserved angular momentum for small oscillations. The sign of $L_z$ determines the nature of orbits, as can be seen in Fig.~\ref{f:A-C-regions-phi1-phi2-square}, with symbol sequences given in Table~\ref{t:static-periodic-symb-seq}. (a) When $L_z \neq 0$, we have elliptical orbits enclosing nonzero area traversed clockwise or anticlockwise. (b) $L_z = 0$ corresponds to straight line trajectories through the origin and are of four sorts. The first three correspond to low-energy pendula where two rotors are always together: (i) $E_1 = 0$ so that $\vf_1 \equiv 0$ (ii) $E_2 = 0$ so that $\vf_2 \equiv 0$ and (iii) $E_1 = E_2$ with $\om_0 (t_2 - t_1) = (2n+1) \pi$ for some integer $n$, so that $\vf_1 + \vf_2 \equiv 0$. (iv) The generic $L_z = 0$ case corresponds to a line through the origin at general inclination so that no two rotors are always together, as in low-energy breathers. 

\vspace{5pt}

\fl {\bf Low-energy nonrotating choreographies.} The (nonrotating) choreographies introduced in \S IV B of Ref.~\cite{gsk-hs-2019} are $3 \tau$ periodic solutions where the three rotors move back and forth on a fixed arc of the circle equally separated in time, leading to successive coincidences. There are two basic types of such choreographies (`clockwise' and `anticlockwise') depending on the order in which the rotors follow each other. At low energies, the clockwise choreography is expressed in terms of the $3\tau$ periodic function $\vf_1(t) = \sqrt{2E_1/3g} \sin(\omega_0 (t-t_1))$ where $3\tau = 2\pi/\om_0$ with $\vf_2(t) = \vf_1(t+\tau)$, so that $\vf_1(t) + \vf_2(t) = -\vf_1(t-\tau)$. We notice that rotors 1 and 2 coincide at $T_n = t_1 + 3n\tau/2$, while 2 and 3 coincide at $T_{n+1}-\tau = T_n + \tau/2$ and 1-3 coincidences occur at $T_n+\tau$ for integer $n$. Therefore, the sequence is $\cdots 312312 \cdots$ (see Table~\ref{t:static-periodic-symb-seq}). The qualifier `clockwise' records the fact that the trajectory may be viewed as an ellipse traversed clockwise in the $\vf_1$-$\vf_2$ square, as in Fig.~\ref{f:A-C-regions-phi1-phi2-square}. Indeed, from (\ref{e:soln-sm-osc}) we find that $E_2 = E_1$ and that the phase difference between $\vf_1$ and $\vf_2$ is $\om_0(t_2-t_1) = -2\pi/3$, so that the ellipse has major axis along the line $\varphi_1 + \varphi_2 = 0$. In the anticlockwise choreography, $\vf_1$ is as before but $\vf_2(t) = \vf_1(t-\tau)$, so that $\vf_1(t)+\vf_2(t) = -\vf_1(t+\tau)$. At low energies, this leads to the same ellipse as the clockwise choreography but traversed anticlockwise (phase difference is $2\pi/3$), with sequence $\cdots 321321\cdots$.

\vspace{5pt}

\fl {\bf Periodic symbol sequence from aperiodic orbit.} While a periodic trajectory necessarily has a periodic symbol sequence, the converse is not always the case. Suppose a trajectory winds around the origin of the $\vf_1$-$\vf_2$ square while maintaining a clockwise or anticlockwise sense. Then, we see from Fig.~\ref{f:A-C-regions-phi1-phi2-square} that the symbol sequence will be periodic ($\cdots 123123 \cdots$ or $\cdots 321321 \cdots$) irrespective of whether the trajectory is periodic. For instance, we find that a small perturbation to the choreography solution at energy $E = 2g$ with ICs $\vf_1(0) = 0$, $\vf_2(0) = 1$ and $p_2(0) = 0$ is a quasiperiodic orbit but has the same periodic sequence $\cdots 123123 \cdots$ as the choreography. 

%-----------------
\subsection{Word-complexity of static and periodic solutions}
\label{s:static-periodic-word-complexity}
%-----------------

\textbf{Static solutions}: Static solutions $\g$ with nonempty $s(\g)$ have symbol sequences of length one and word-complexity $C_n = \del_{n,0}$ for $n \geq 0$.

\vspace{4pt}

\fl \textbf{Periodic solutions}: A periodic trajectory $\g$ has a periodic symbol sequence. Consider a $p$-periodic sequence: $x_1 x_2 \cdots x_p x_{p+1} \cdots$, with $x_{p+1} = x_1, x_{p+2} = x_2$, etc. This means the smallest $j \geq 1$ for which $x_k = x_{j+k}$ for all $k = 1,2,3, \dots$ is $j = p$. For the periodic  pendula, breathers and small oscillations around G discussed in \S \ref{s:symb-seq-static-periodic} and listed in Table~\ref{t:static-periodic-symb-seq}, we find that the word-complexity is equal to the period, $C_n = p$, so that their topological entropy vanishes. Note that shifts of the six small-oscillation symbol sequences give rise to thirteen distinct sequences. They illustrate a proposition of Morse and Hedlund \cite{morse-hedlund-1940,bruin-book-2022}: if the word-complexity of a subshift $(X, \tau)$ satisfies $C_n \le n$ for some $n$, then $(X, \tau)$ consists of finitely many periodic sequences. Let $X$ denote the space of symbol sequences arising from small oscillations around G. In this case, it turns out that $C_n = 13$ for all $n \ge 2$, implying that $X$ must consist of finitely many periodic sequences (thirteen in this case, with $p = 1,2,3$).

%-----------------
\section{Rotor coincidences must be crossings}
\label{s:coinc-and-cross}
%-----------------

Although they are identical particles, rotors are distinguishable since rotor angles arise as superconducting phases on distinct metallic segments in a chain of coupled Josephson junctions (see Appendix A of Ref.~\cite{gsk-ay-2023}). Viewed as points on a circle, when a pair of rotors meet, must they pass through each other (i.e., cross) or can they reverse directions and turn back? Our upcoming results on pair and triple coincidences imply that any rotor coincidence must involve a reversal of rotor order (A to C or C to A) provided the order is initially defined (i.e., no two rotors are always together). Thus, rotors must pass through each other when they meet. It follows that we may infer the CA symbol sequence from that in the 0123 alphabet provided the initial order is well-defined and known. Note that if two rotors have a common position and velocity at one instant, then they are subject to the same force and should continue to be together. In fact, by time-reversibility, they would have been together in the past as well, forming a molecule. In this case, the rotor order is not well-defined.

%-----------------
\subsection{Pairwise coincidences}
\label{s:pair-coinc-cross}
%-----------------

We argue that an isolated coincidence of two rotors must involve a crossing irrespective of the state of the third rotor. Imagine rotor 1 catching up with rotor 2. The case where they are approaching each other head-on is related by a change of frame. In order for them to meet, 1 must have a higher speed than 2 just before the meeting. However, if 1 continues to have a higher speed at the meeting point, then it would overtake 2, resulting in a crossing. So, to have a meeting but avoid a crossing, 1 must decelerate and reach the same speed as 2 at the meeting point. However, in this case, the two rotors would have the same position and velocity at the meeting instant implying that their trajectories would coincide at all times. In other words, a coincidence must either be an eternal meeting or a crossing. While the pendulum solutions give examples of the former, the latter are to be found in breathers as well as in choreographies. We note that these qualitative observations are independent of the position and velocity of the third rotor: the force it exerts depends only on the relative angles, which are the same for rotors 1 and 2 when they meet. 

Another way of seeing that an instantaneous coincidence of two rotors (say, 1 and 2) must involve a crossing is to examine the EOM (\ref{e:phi-eom}) for $\vf_1 = \tht_1 - \tht_2$ in the vicinity of a 1-2 coincidence, where $\vf_1$ is small:
	\beq
	\ddot \vf_1 = -(g/mr^2) (2 \sin \vf_1 - \sin \vf_2 + \sin(\vf_1 + \vf_2)) 
	\approx - (g/mr^2) (2 + \cos \vf_2) \vf_1.
	\label{e:pair-coinc}
	\eeq
This is the equation for simple harmonic motion of $\vf_1$ due to a $\vf_2$-dependent linear restoring force, since $(g/m r^2)(2 + \cos \vf_2) > 0$. Moreover, when $\vf_1 \to 0$, $\vf_2$ approaches a limiting value. As is well known, the oscillator cannot reverse direction at an equilibrium point. This means the sign of $\vf_1$ must change at a 1-2 coincidence implying that the rotors must cross irrespective of the state of the third rotor.

%-----------------
\subsection{Triple coincidences}
\label{s:triple-coinc-cross}
%-----------------

Does a triple coincidence imply that the rotor order must change from A to C or vice versa? In \S \ref{s:symb-seq-static-periodic}, we saw that in isosceles breathers, a triple coincidence resulted in order reversal. We now argue that this is generally true. In fact, a triple coincidence may be viewed as involving three simultaneous pairwise coincidences that together reverse the rotor order. We will first present an argument in the linear approximation and then proceed to a more general demonstration.

A triple coincidence occurs when $\varphi_1 = \varphi_2 = 0$ corresponding to the point G at the origin of the $\vf_1$-$\vf_2$ square of Fig.~\ref{f:A-C-regions-phi1-phi2-square}. Near G, the linearized EOM  $mr^2 \ddot \varphi_{1,2} = -3g \varphi_{1,2}$ are the same as those for small oscillations presented in \S \ref{s:symb-seq-static-periodic} with solutions given in Eq.~(\ref{e:soln-sm-osc}). Moreover, the conserved angular momentum $L_z$ (\ref{e:ang-mom-sm-osc-and-psi-dot}) vanishes at G. So, a trajectory with a triple coincidence must have $L_z \equiv 0$ near G in the linear approximation. From (\ref{e:ang-mom-sm-osc-and-psi-dot}), $L_z$ vanishes when $E_1 = 0$ or $E_2 = 0$ or $\om_0 (t_2 - t_1) = n \pi$ for $n$ an integer. These comprise all straight lines passing through the origin G. Referring to Fig.~\ref{f:A-C-regions-phi1-phi2-square}, this implies that in the linear approximation, order (if defined) must reverse at a triple coincidence. The problem with extending this argument beyond the linear approximation is that there could be a trajectory that curves for instance from a C cell to another C cell but has a tangent at G that lies along the boundary between A and C. Such a trajectory would not involve a reversal of rotor order at a triple coincidence, but the linear approximation would mistakenly say that the order was not even defined. We now give an argument that does not use the linear approximation.
	
Suppose rotor 1 is catching up with 2, and rotor 3 is approaching them head-on, with initial order 1-2-3 clockwise. Suppose they undergo an order-preserving triple coincidence with final order being 3-1-2 clockwise. Prior to the coincidence, 1 must be moving faster than 2. After the coincidence, 1 must be slower than 2 for the latter to be ahead of 1. This means 1 had to be decelerating around the time of the coincidence. However, since 2 and 3 were initially in front of 1, they would attract and accelerate it, leading to a contradiction. Alternatively, as argued for pair coincidences, at the meeting point, 1 and 2 must have the same velocity, which means they would have to be coincident at all times. Therefore, our assumption of order-preservation is wrong. Another way of preserving order is for 3 to end up between 1 and 2 (order 2-3-1). This too cannot happen since 2 and 3 would have to come to rest at the meeting point whence they would have had to be together prior to the coincidence. Therefore, order, if defined, must reverse at a triple coincidence. Also, if two rotors are stuck together, they would remain so during a triple coincidence: this corresponds to trajectories along boundaries between the A and C regions.

%-----------------
\section{Ultra-high-energy orbits: symbol sequences, their grammar and word statistics}
\label{s:UHE-ss}
%-----------------

This section is devoted to an ultra-high-energy (UHE) limit where rotor kinetic energy dominates over interactions. The resulting trajectories are either periodic or quasiperiodic. We examine structural features of their symbol sequences and find that their word-complexities either saturate or grow linearly. Several grammar rules are identified but it is argued that UHE sequences cannot be modeled using a topological Markov chain (subshift of finite type, defined by a finite list of forbidden words). Nevertheless, we are able to explain the linear growth of word-complexity for UHE quasiperiodic sequences through an analysis of right-special words and a fairly natural connection to Sturmian sequences.

%-----------------
\subsection{Structure of trajectories and their symbol sequences}
\label{s:structure-traj-sym-seq-UHE}
%-----------------

\begin{figure}[t]
	\centering
	\begin{subfigure}[b]{0.40\textwidth}
		\centering
		\includegraphics[width=\textwidth]{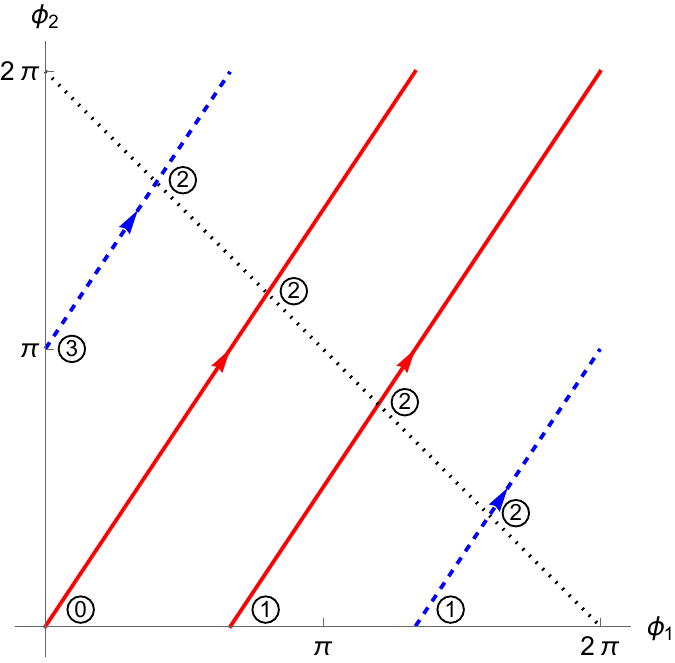}
		\caption{$\sig = 3/2$}
		\label{f:highE-sig-3/2}
	\end{subfigure}
	\hspace{0.025\textwidth}
	\begin{subfigure}[b]{0.40\textwidth}
		\centering
		\includegraphics[width=\textwidth]{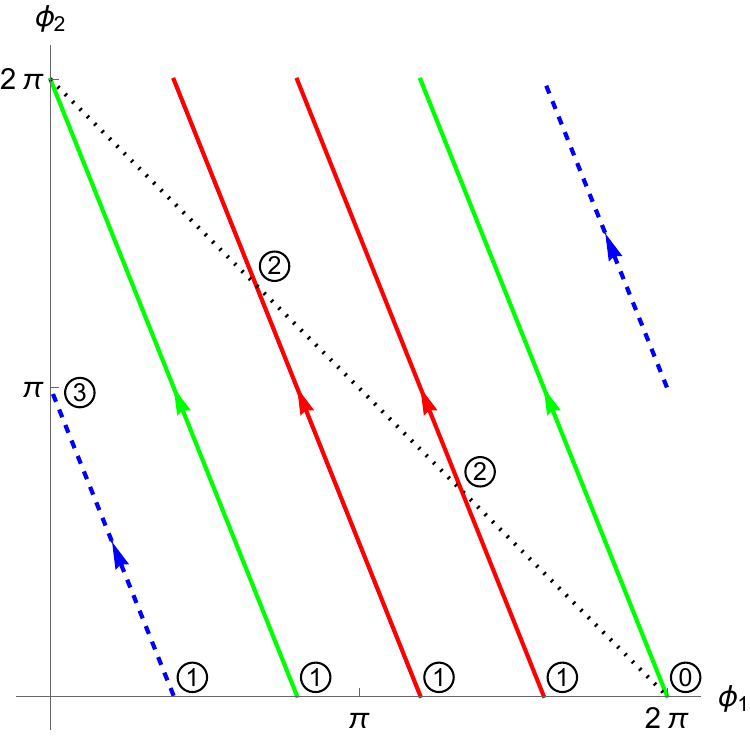}
		\caption{$\sig = -5/2$}
		\label{f:highE-sig--5/2}
	\end{subfigure}
	\caption{\small Two periodic UHE trajectories on the $[0,2\pi)^2$ fundamental domain of the $\vf_1$-$\vf_2$ configuration space: (a) $\sig = 3/2$ and (b) $\sig = -5/2$. Arrows indicate the direction of time assuming $\om_2 > 0$ with trajectories beginning at the origin. The solid red lines are unbroken lines that cross the dotted black 3-1 coincidence curve. The solid green lines are unbroken lines that do not involve 3-1 pair coincidences. Dashed blue lines are broken lines. The symbols at rotor coincidences are enclosed in circles. The repeating blocks in the symbol sequences for (a) 02 1232 12 and (b) 0 12 13 12 1 are related by 1-2 rotor exchange which also relates $\sig = 3/2$ to $\sig = -5/3$ via $\sig \to -(1 + \sig)$.}
	\label{f:highE-plot-1}
\end{figure}

\begin{figure}[ht]
	\centering
	\begin{subfigure}{0.3\textwidth}
		\centering
		\includegraphics[width=\textwidth]{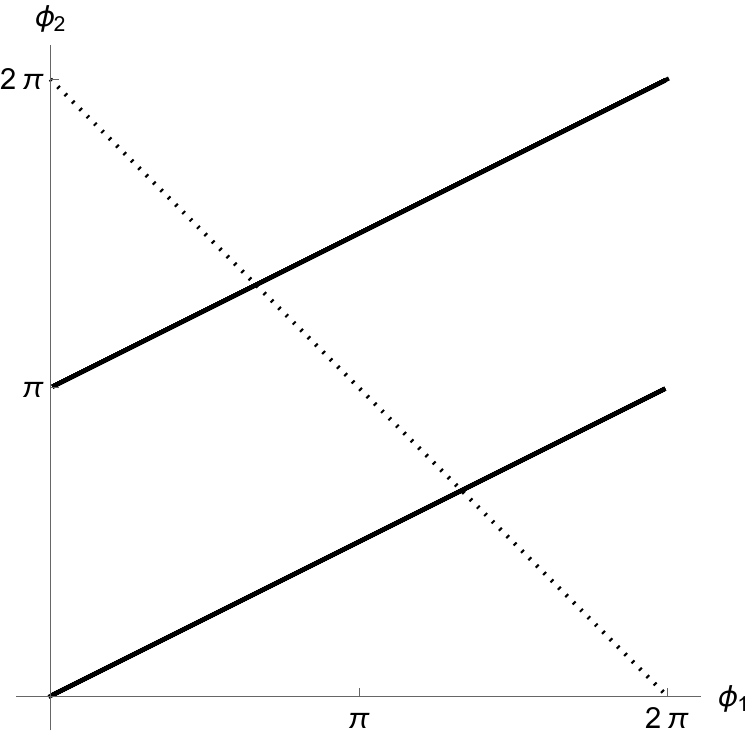}
		\caption{$\sig = 1/2$}
		\label{f:highE-sig-1/2}
	\end{subfigure}
	\begin{subfigure}{0.3\textwidth}
		\centering
		\includegraphics[width=\textwidth]{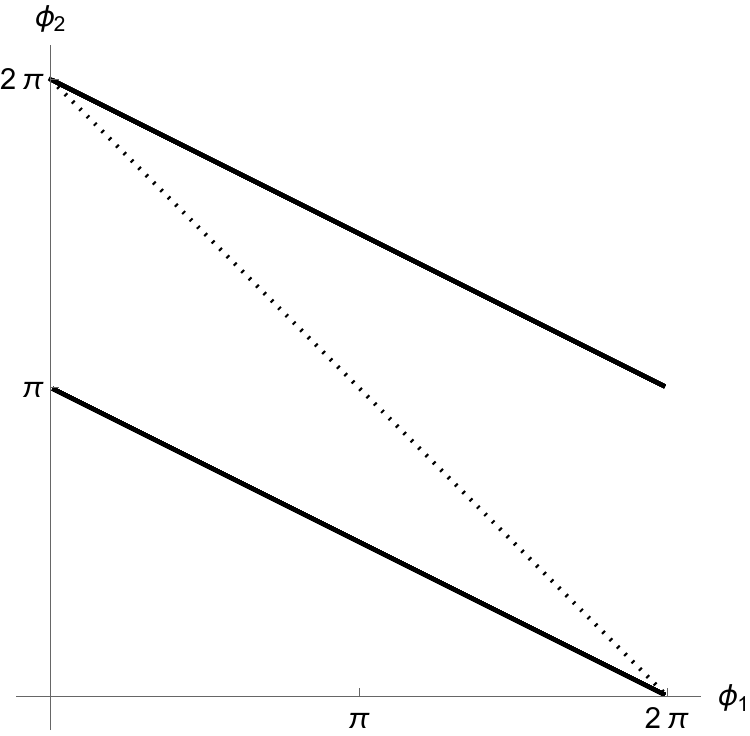}
		\caption{$\sig = -1/2$}
		\label{f:highE-sig--1/2}
	\end{subfigure}
	\begin{subfigure}{0.3\textwidth}
		\centering
		\includegraphics[width=\textwidth]{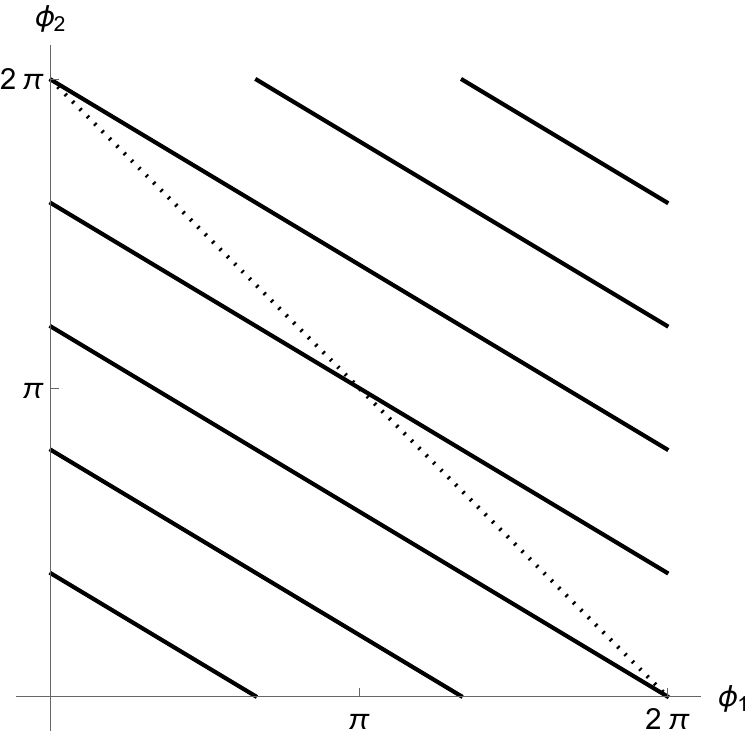}
		\caption{$\sig = -3/5$}
		\label{f:highE-sig--3/5}
	\end{subfigure}
	
	\begin{subfigure}{0.3\textwidth}
		\centering
		\includegraphics[width=\textwidth]{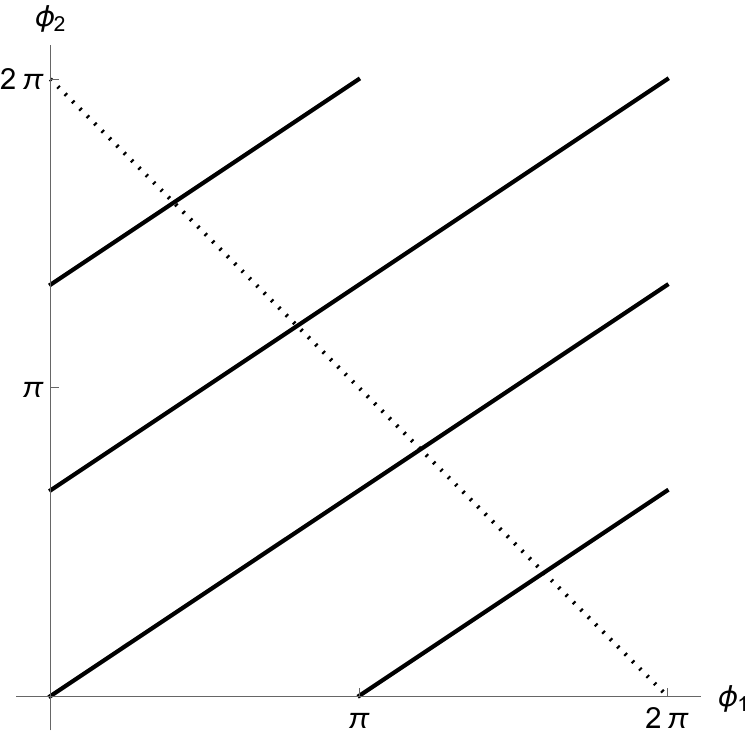}
		\caption{$\sig = 2/3$}
		\label{f:highE-sig-2/3}
	\end{subfigure}
	\begin{subfigure}{0.3\textwidth}
		\centering
		\includegraphics[width=\textwidth]{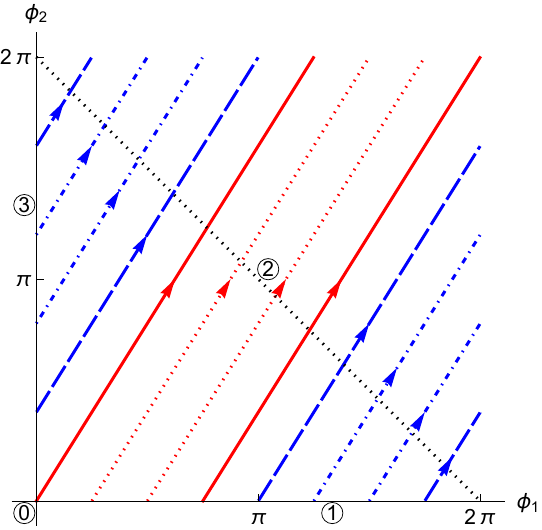}
		\caption{$\sig = 8/5 = 1.6$}
		\label{f:highE-sig-8/5}
	\end{subfigure}
	\begin{subfigure}{0.3\textwidth}
		\centering
		\includegraphics[width=\textwidth]{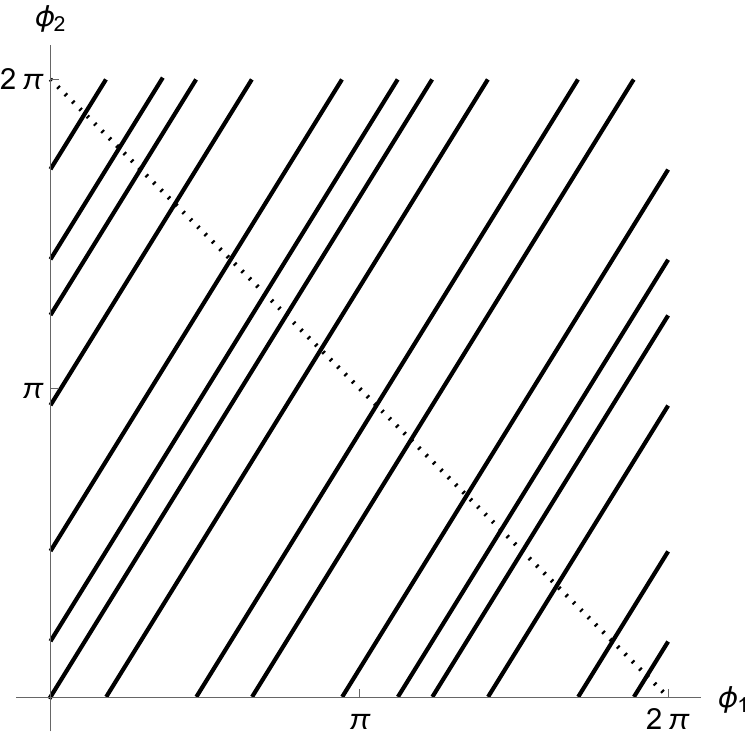}
		\caption{$\sig = (1+\sqrt{5})/2 = 1.618 \dots$}
		\label{f:highE-sig-golden-ratio}
	\end{subfigure}
\caption{\small $\vf_1$-$\vf_2$ configuration space showing some ultra-high-energy trajectories: (a) $\sig = 1/2$, (b) $\sig = -1/2$, (c) $\sig = -3/5$, (d) $\sig = 2/3$ and (e) $\sig = 8/5 = 1.6$, plotted over one period, and (f) $\sig = (1+\sqrt{5})/2 = 1.618 \dots$ (golden ratio), plotted up to 10 cycles of $\vf_2$. The symbol sequences for (e) and (f) are the same up to 8 $\vf_2$ cycles but differ thereafter. In (e), for $\sig = 8/5$, the two consecutive unbroken lines (solid red), isolated unbroken lines (dotted red), two consecutive broken lines (dashed blue) and isolated broken lines (dot-dashed blue) are indicated.}
	\label{f:highE-plot-2}
\end{figure}

There are many different high-energy limits one could consider. For instance, pendulum solutions exist at all relative energies and we could consider trajectories in the vicinity of high energy pendula. Rather than do this, we will consider a very simple `ultra-high-energy' (UHE) limit where $E \gg g$ and the kinetic terms dominate the interactions. The EOM (\ref{e:phi-eom}) reduce to $\ddot \vf_{1,2} \equiv 0$ so that the relative angles evolve linearly:
	\beq
	\vf_1 (t) = \om_1 t + \al_1 \quad \text{and} \quad \vf_2 (t) = \om_2 t + \al_2,
	\label{e:uhe-linear-evol-rel-angles}
	\eeq
where $\om_{1,2}$ are individual angular frequencies and $\al_{1,2}$ initial angles. Such a trajectory is represented by a straight line on the $\vf_1$-$\vf_2$ square with opposite sides identified, as in Figs.~\ref{f:highE-plot-1} and \ref{f:highE-plot-2}. The time periods of the relative angles $\vf_1$ and $\vf_2$ are $T_{1,2} = 2\pi/|\om_{1,2}|$. The orbit is periodic if $T_1$ and $T_2$ have a common integer multiple, i.e., when the slope $\sig = \om_2/\om_1$ is rational. If $\sig$ is irrational, the trajectory is quasiperiodic and comes arbitrarily close to any point on the torus after a sufficiently long time. 

We now suppose that $\al_1$ and $\al_2$ are both integer multiples of $2\pi$, so that these straight line trajectories $\g_\sig$ begin at the origin; other trajectories can be dealt with in a similar manner. If $\om_1, \om_2 > 0$, then the arrow of time is Southwest to Northeast, while it is SE--NW if $\om_1 < 0 < \om_2$. Reversal of signs of both $\om_1$ and $\om_2$ leads to the time-reversed trajectory. Where convenient, we will assume that $\om_2 > 0$. 

Leaving aside trajectories with eternal coincidences (which happen when $\sig = 0, -1, \pm \infty$), rotors undergo a succession of crossings resulting in a CA symbol sequence of the form $\cdots$CACA$\cdots$. The sequence $s(\g_\sig)$ of such a trajectory in the 0123 alphabet is much more interesting. It is obtained by recording the symbols 0, 1, 2 or 3 when the trajectory passes through the origin or intersects the $\vf_1$ axis (floor), diagonal 3-1 coincidence curve or the $\vf_2$ axis (wall). Notably, the manner in which these UHE symbol sequences are generated bears a resemblance to the construction of Sturmian sequences as `cutting sequences' of straight lines with irrational slope on a square grid. This suggests a connection between our UHE quasiperiodic symbol sequences and Sturmian sequences, which we develop in \S \ref{s:sturmian}. In the rest of this section, we discuss some properties and structural features of these UHE trajectories and their symbol sequences.

\vspace{5pt}

{\fl \bf Rational slope.} Suppose $\sig = p/q$, where $p$ and $q$ are coprime integers. We may write $\om_1 = q/\tau$ and $\om_2 = p/\tau$ for some positive real time scale $\tau$. Then, the time period is 
	\beqs
	T &=& \lcm(T_1, T_2) = 2\pi \tau \: \lcm(1/|q|, 1/|p|) = 2 \pi \tau
	= 2\pi p/\om_2 = 2\pi q/\om_1.
	\label{e:time-period-rational-slope-uhe}
	\eeqs
In one time period $T$, $\vf_1$ goes through $|q|$ cycles while $\vf_2$ goes through $|p|$ cycles. The symbol sequence of a periodic orbit starting at the origin is the same as that of the time-reversed orbit (see Figs.~\ref{f:highE-plot-1} and \ref{f:highE-plot-2}): ceiling and floor and left and right walls get exchanged, i.e., the abscissae and ordinates of the points of intersection are replaced by their complements modulo $2\pi$: $(\vf_1, \vf_2) \mapsto (2\pi - \vf_1, 2\pi - \vf_2)$. Thus, $s(\g_{p/q})$ is a palindrome between successive 0s: $s(\g_\sig) = .0abc \cdots ghg \cdots cba0 abc \cdots ghg \cdots cba0 \cdots$. The examples $s(\g_1) = s(\g_{-1}) = .02020 \cdots$, $s(\g_0) = .0101 \cdots$, $s(\g_{\pm \infty}) = .0303 \cdots$ and those in Table~\ref{t:uhe-rat-slope} illustrate this. We will soon give a formula for the period $N_\sig$ of the symbol sequence $s(\g_\sig)$.

\begin{table}[h!]
	\begin{center}
		\footnotesize
		\begin{tabular}{|c|c|c|}
			\hline
			\textbf{Slope $\boldsymbol{\sig}$} & \textbf{Repeating block} & \textbf{Period} $\boldsymbol{N_\sig}$\\
			\hline
			2 & 02 12 & 4 \\
			1/2 & 02 32 & 4 \\
			-1/2 & 03 & 2 \\
			3/2 & 02 1232 12 & 8 \\
			-5/2 & 0 12 13 12 1 & 8 \\
			-3/5 & 0 31 32 31 3 & 8 \\
			2/3 & 02 3212 32 & 8 \\ 
			8/5 & 02 1232 12 1232 1232 12 1232 12 & 24 \\
			\hline
		\end{tabular}
	\end{center}
\caption{\small Repeating (palindromic) block in symbol sequences of some rational slope UHE trajectories. Spaces separate successive $\vf_1$ or $\vf_2$ cycles according as $|\sig| < 1$ or $|\sig| > 1$.}
	\label{t:uhe-rat-slope}
\end{table}

\fl {\bf Irrational slope.} If $\sig$ is irrational, then the trajectory is quasiperiodic. The symbol sequence is not periodic and does not display the palindrome property. For example, the trajectory for $\sig = \vf = (1+\sqrt{5})/2 = 1.618 \dots$ is shown in Fig.~\ref{f:highE-sig-golden-ratio} and we may infer that
	\beq
	s(\g_\vf) = \text{.02 1232 12 1232 1232 12 1232 12 1232 1232} \cdots.
	\eeq
We can get as many digits as we want using a suitable rational approximation to $\vf$. For instance, from Fig.~\ref{f:highE-sig-8/5} and Table \ref{t:uhe-rat-slope} we see that $s(\g_{8/5})$ agrees with the first 24 letters of $s(\g_\vf)$.

\vspace{5pt}

{\fl \bf Using rotor exchanges to restrict to $\boldsymbol{\sig > 1}$.} We see from Fig.~\ref{f:highE-plot-1} that the manner in which the trajectories intersect the  `diagonal' 3-1 coincidence line is different for $\sig > 0$ and $\sig < 0$. However, using rotor exchanges introduced in \S \ref{s:rot-ord-rot-coincidence-sym-seq}, one may obtain the symbol sequence for a trajectory with $\sig < 1$ from that for a trajectory with $\sig > 1$. To understand the effect of exchanges, suppose we begin with the three rotors at angular positions $(\tht_1, \tht_2, \tht_3)$ and then exchange rotors 1 and 3. The angular positions of the rotors are now $(\tl \tht_1, \tl \tht_2, \tl \tht_3) = (\tht_3, \tht_2, \tht_1)$. Under this exchange, the relative angles transform as $\vf_1 \to \tl \vf_1 = - \vf_2$ and $\vf_2 \to \tl \vf_2 = - \vf_1$. Thus, the original slope $\sig = \dot \vf_2/\dot \vf_1$ is mapped to $\tl \sig = \dot {\tl \vf}_2/\dot{\tl \vf}_1 = 1/\sig$. The effect of other rotor permutations is given in Table~\ref{t:rotor-exchanges} and may be used to restrict to $\sig > 1$ in the following manner:

\begin{itemize}
	\item[(1)] The sequence for a trajectory with $0 < \sig \leq 1$ is obtained from one with $\tl \sig = 1/\sig \geq 1$ by a $3 \leftrightarrow 1$ symbol exchange. For example, $s(\g_{2/3})$ = .02 3212 32 $\cdots$ of Fig.~\ref{f:highE-sig-2/3} is deduced from $s(\g_{3/2})$ = .02 1232 12 $\cdots$.
	\item[(2)] The sequences for $\sig < -1$ are got from those with $\tl \sig = -(1+\sig) > 0$ by exchanging symbols 1 and 2. For example, $s(\g_{-5/2})$ = .0 12 13 12 1 $\cdots$ of Fig.~\ref{f:highE-sig--5/2} is obtained from $s(\g_{3/2})$.
	\item[(3)] Finally, slopes $-1 < \sig \leq 0$ may be obtained from positive slopes $\tl \sig = -\sig/(1 + \sig)$ by $2 \leftrightarrow 3$ exchange. For example, $s(\g_{-3/5})$ = .0 31 32 31 3 $\cdots$ of Fig.~\ref{f:highE-sig--3/5} is obtained from $s(\g_{3/2})$.
\end{itemize}

\begin{table}[ht]
	\begin{center}
		\footnotesize
		\begin{tabular}{|c|c|c|c|c|}
			\hline
			Permutations & Action on $(\tht_1, \tht_2, \tht_3)$ & Action on $(\vf_1, \vf_2)$ & Action on $\sig$ & Action on CA, 0123\\
			\hline
			$e$ & $(\tht_1, \tht_2, \tht_3)$ & $(\vf_1, \vf_2)$ & $\sig$ & CA, 0123\\
			$\tau_{12}$ & $(\tht_2, \tht_1, \tht_3)$ & $(-\vf_1, \vf_1 + \vf_2)$ & $-(1 + \sig)$ & AC, 0213\\
			$\tau_{23}$ & $(\tht_1, \tht_3, \tht_2)$ & $(\vf_1 + \vf_2, -\vf_2)$ & $-\sig/(1 + \sig)$ & AC, 0132\\
			$\tau_{31}$ & $(\tht_3, \tht_2, \tht_1)$ & $(-\vf_2, -\vf_1)$ & $1/\sig$ & AC, 0321\\
			$\sig_{123} = \tau_{23} \tau_{12}$ & $(\tht_2, \tht_3, \tht_1)$ & $(\vf_2, -\vf_1 -\vf_2)$ & $-(1 + 1/\sig)$ & CA, 0312\\
			$\sig_{132} = \tau_{12} \tau_{23}$ & $(\tht_3, \tht_1, \tht_2)$ & $(-\vf_1 -\vf_2, \vf_1)$ & $-1/(1 + \sig)$ & CA, 0231\\
			\hline
		\end{tabular}
	\end{center}
\caption{\small Action of rotor exchanges ($\tau_{ij}$) and cyclic permutations ($\sig_{klm}$) on rotor angles, relative angles, slope $\sig$ of UHE trajectories and symbols (see also Table~I of Ref.~\cite{gsk-hs-2024}).}
	\label{t:rotor-exchanges}
\end{table}

% Note that the cyclic permutations can be expressed as compositions of two rotor exchanges:   $\sig_{123} = \tau_{23} \tau_{12} = \tau_{31} \tau_{23} = \tau_{12} \tau_{31}$ and $\sig_{132} = \tau_{31} \tau_{12} = \tau_{12} \tau_{23} = \tau_{23} \tau_{31}$. (An extension of Table~I of \cite{gsk-hs-2024}.)

{\fl \bf Some features of trajectories with slope $\boldsymbol{\sig > 1}$.} We now focus on cases where $\sig = \om_2/\om_1 > 1$. For definiteness, let $\om_2 > \om_1 > 0$ so that lines `climb' faster than they `run' and the trajectory hits the ceiling more often than the wall, as in Fig.~\ref{f:highE-sig-3/2}. In terms of rotors, the angle $\vf_1$ between rotors 1 and 2 grows slower than the angle $\vf_2$ between rotors 2 and 3. To understand some structural features of the corresponding symbol sequence, we introduce the notions of unbroken and broken full lines. We call one cycle of $\vf_2$ (from $0$ to $2\pi$) a full line: it extends from floor to ceiling. A full line is {\it unbroken} (solid red lines in Fig.~\ref{f:highE-sig-3/2}) if it extends from floor to ceiling without meeting a wall (except perhaps at (0,0)). An unbroken full line can correspond to three possible words: 021, 121 and 120. A full line is {\it broken} (dashed blue lines in Fig.~\ref{f:highE-sig-3/2}) if it hits a wall before reaching the ceiling; it corresponds to the word 12321. Breaking happens when $\vf_1$ reaches $2\pi$ before $\vf_2$ does. Broken lines come in two pieces and cannot pass through the origin as $\sig > 1$. In particular, the first full line of any orbit that begins at the origin must be unbroken, as must the last full line in one period of a periodic orbit that begins at the origin. 

We now deduce some structural features of UHE sequences with $\sig > 1$ bearing in mind the period 8 example $s(\g_{3/2}) = $ .02 1232 12 02 1232 12 $\cdots$ shown in Fig.~\ref{f:highE-sig-3/2}. We begin by noting that every full line intersects the 3-1 coincidence diagonal. Unbroken lines cross the diagonal once while broken lines cross it twice. Consequently, there are alternating 2s in the symbol sequence. In fact, an unbroken line contributes 02 (origin-diagonal) if it is the first line of the trajectory and 12 (floor-diagonal) otherwise. Since the last line of a periodic trajectory must be unbroken, we deduce that the repeating block of a periodic sequence must be of the form $.02 \cdots \un{\;\;} 2 \un{\;\;} 2 \un{\;\;}$  $\cdots 12$. Here, the blanks are placeholders for 1s and 3s. On the other hand, a broken line contributes four letters: 1232 (floor-diagonal-wall-diagonal). In particular, any $\sig > 1$ symbol sequence will have more 1s than 3s. For example, the $\sig = 3/2$ periodic trajectory consists of a full line (02) followed by a broken line (1232) and another full line (12).

\vspace{5pt}

{\fl \bf Period of symbol sequence for rational $\boldsymbol{\sig > 1}$.} Since we know the words associated with unbroken and broken lines, we will count how many such lines there are in a UHE periodic trajectory with rational $\sig = p/q > 1$ and thereby find the length of the repeating block in its symbol sequence. From (\ref{e:time-period-rational-slope-uhe}), there must be $p$ full lines per period as there are $p$ $\vf_2$ cycles in a period. Breaking happens when $\vf_1$ alone reaches $2\pi$. Since there are $q$ $\vf_1$ cycles in a period, we get $q-1$ broken lines (the last line ends at $(2\pi, 2\pi)$ and is regarded as unbroken). Subtracting, the number of unbroken lines is $p-q+1$. Recalling that an unbroken line contributes two letters (02 or 12) while a broken line contributes four (1232), the number of letters per period is
	\beq
	N_{\sig > 1} = 2(p-q+1) + 4(q-1) = 2(p + q -1).
	\label{e:num-lett-1-period-sig-g-1}
	\eeq
This period may be partitioned into the number of times each symbol occurs: there is only one 0 (at the beginning), there are $(p-1)$ 1s (one from each full line except the first one), $(1/2) N_{\sig>1} = (p+q-1)$ 2s (due to the alternating 2s) and $(q-1)$ 3s (one from each broken line). This gives the frequencies of 1-letter words in UHE periodic orbits with $\sig > 1$ (see \S \ref{s:word-freq-uhe-quasiper} for more on word frequencies).

On applying the rotor exchange transformations of Table \ref{t:rotor-exchanges} to (\ref{e:num-lett-1-period-sig-g-1}), we find the period for any nonzero rational $\sig = p/q$:
	\beq
	N_{p/q} =
	\begin{cases}
		2(|p| + |q| - 1) & \text{for} \quad \sig > 0, \\
		2(\max\{|p|,|q|\} - 1) & \text{for} \quad \sig <0, \sig \ne -1.
	\end{cases}
	\label{e:num-lett-1-period}
	\eeq
This formula fails for trajectories with eternal coincidences, which occur when $\sig = 0, -1, \pm \infty$ and have period $N = 2$.

\vspace{5pt}

\fl {\bf Consecutive unbroken and broken lines.} We now find the possible numbers $N_{ub}$ and $N_b$ of consecutive unbroken and broken lines in a UHE orbit of slope $\sig > 1$ starting at the origin. These formulas will help formulate grammar rules for the corresponding symbol sequences in \S \ref{s:gram-rule-uhe} and estimate their word-complexity in \S \ref{s:asymp-argue-Cn-formula}.

\fl {\bf Consecutive unbroken lines.} If $\sig > 1$ is an integer, then all lines are unbroken, leading to an infinite number of consecutive unbroken lines. Next, suppose $\sig$ is irrational, so that the trajectory never returns to the origin. In this case, we will show that $N_{ub} = \lfloor \sig \rfloor - 1$ or $\lfloor \sig \rfloor$ by finding the minimal and maximal values of $N_{ub}$. Here, the floor $\lfloor x \rfloor$ is the greatest integer $\leq x$. We begin by observing that an unbroken line corresponds to an increase in $\vf_1$ by $2\pi/\sig$. It follows that the maximal number of consecutive unbroken lines is given by $\lfloor 2\pi/ (2 \pi/\sig) \rfloor = \lfloor \sig \rfloor$. In particular, the first $\lfloor \sig \rfloor$ lines of the trajectory are unbroken. To obtain the lower bound, we note that the first in a sequence of unbroken lines must start from the floor somewhere in the interval $0 \leq \vf_1 < 2\pi/\sig$ leading to a `loss' of at most $2\pi/\sig$ of the $\vf_1$ axis. Thus, there must be at least $\lfloor (2\pi - 2\pi/\sig)/(2\pi/\sig) \rfloor = \lfloor \sig \rfloor - 1$ unbroken lines in the sequence. The case of rational but nonintegral $\sig$ is similar, except that the trajectory returns to the origin after one period. Thus, there can be $\lfloor \sig \rfloor - 1, \lfloor \sig \rfloor$ or $2 \lfloor \sig \rfloor$ consecutive unbroken lines. The third possibility arises when the last $\lfloor \sig \rfloor$ unbroken lines of one cycle are followed by the first $\lfloor \sig \rfloor$ such lines of the next cycle. For example, for $\sig = 8/5$, we can have $N_{ub} = 1$ or $2$, as we see from Fig.~\ref{f:highE-sig-8/5}.

\vspace{5pt}

\fl {\bf Consecutive broken lines.} We note that a broken line arises each time the trajectory hits a wall (see Fig.~\ref{f:highE-sig-8/5}). Two consecutive broken lines arise if there is only one encounter with the ceiling between two successive wall crossings, which corresponds to one $\vf_1$ cycle. Let $N_b$ denote the possible numbers of consecutive broken lines in a $\sig > 1$ orbit that starts at the origin. For example, (a) $N_b = 1$ if $\sig = 3/2$, (b) $N_b = 2$ if $\sig = 4/3$ and (c) $N_b = 1$ or $2$ if $\sig = 8/5$ (see Figs.~\ref{f:highE-plot-1} and \ref{f:highE-plot-2}). For integer $\sig > 1$, there are no broken lines, as such an orbit is periodic and never hits a wall except at the origin. For nonintegral $\sig > 2$, broken lines must be isolated: $N_b = 1$. This is because $\vf_2$ would complete at least two cycles by the time $\vf_1$ completes one, leading to at least one unbroken line between any two broken lines. 

For $1 < \sig < 2$, we now show that $N_b$ can take only two values: $M - 1 \le N_b \le M$, where $M - 1 = \lfloor 1/(\sig - 1) \rfloor_s$ and $M = \lfloor \sig / (\sig - 1) \rfloor_s$. Here, the strict floor $\lfloor x \rfloor_s$ is the greatest integer strictly less than $x$. 

Suppose we start on the first of a sequence of broken lines at a wall crossing. The next wall crossing happens after one $\vf_1$ cycle, which corresponds to a $2 \pi \sig$ increment in $\vf_2$. Modulo $2\pi$, this is $\D \vf_2 = 2\pi (\sig - 1)$, which is the height the trajectory climbs along the wall between two such wall crossings. The appearance of an unbroken line results in a drop in this height. Thus, the maximum number of consecutive broken lines is the number of successive wall crossings before the appearance of an unbroken line, which in turn is one more than the largest number of such $\vf_2$ increments before a drop. This is given by the largest integer less than $1 + 2\pi/\D \vf_2 = \sig/(\sig - 1)$. Hence, $N_b \leq M$. 

To arrive at the lower bound on $N_b$, we observe that the first in a sequence of broken lines must cross the wall at a height $\leq \D \vf_2$. The above argument then implies that the minimum number of consecutive broken lines is the largest integer less than $1 + (2\pi - \D \vf_2)/\D \vf_2 = 1/(\sig - 1)$. Thus, $N_b \geq M-1$. For irrational $\sig$, $N_b$ takes both the values $M-1$ and $M$. For rational $\sig = p/q$ with $p$ and $q$ coprime and positive, the larger value $M$ is achieved only if $p < (M q - 1)/(M-1)$. To see this, suppose the first in a sequence of $M$ consecutive broken lines begins at height $\vf_2^0$ on a wall. To accommodate these $M$ wall crossings in the upper portion of the wall beginning at $\vf_2^0$, we must have $1 + (2 \pi - \vf_2^0)/\D \vf_2 > M$. Next we get a lower bound on $\vf_2^0$. Since the periodic orbit began at the origin, from (\ref{e:time-period-rational-slope-uhe}), there must be $q$ $\vf_1$ cycles in a period and a total of $q-1$ wall crossings (aside from the origin) with height spacing of $2\pi/q$. It follows that $\vf_2^0 \geq 2\pi/q$. Combining the two inequalities, we get the advertised upper bound on $p$. For instance, when $\sig = 3/2$, this inequality is violated so that broken lines must be isolated. Interestingly, it may be shown that the larger value of $N_b$ is always achieved if the periodic trajectory does not pass through the origin.

% See extra notes file for proof of statement that larger value of $N_b$ is always achieved if the periodic trajectory does not pass through the origin.

%-----------------
\subsection{Word statistics of periodic trajectories}
\label{s:uhe-periodic}
%-----------------

We now examine the word statistics of UHE periodic orbits ($\sig$ rational) that start at the origin. It is instructive to look at some examples from Table \ref{t:uhe-rat-slope}. Notice that (a) $s(\g_2) = 02120 \cdots$ has period $p=4$ and word-complexity $C_1 = 3$ and $C_n = 4$ for $n \ge 2$, (b) $s(\g_{3/2}) = 021232120 \cdots$ has period 8 with $C_1 = 4$, $C_2 = 6$, $C_3 = 7$ and $C_n = 8$ for $n \ge 4$, and (c) $s(\g_{8/5})$ is 24-periodic and its word-complexity grows as $2n+2$ for $n \le 8$, as $n + 10$ for $8 \leq n \leq 14$ and then plateaus at $C_n = 24$. Additional examples indicate that this pattern holds for other rational $\sig$: $C_n$ first grows with slope $2n$, then with slope $n$ and finally saturates at $C_n = p$ at some $n \le p$. In particular, we find that for UHE periodic trajectories, $C_n = p$ for $n \ge p$. Let us now show why this is the case. Suppose the trajectory passes through the origin and has a $p$-periodic symbol sequence. The repeating block may be taken to be of the form $x_1 x_2 \cdots x_p$ where $x_1 = 0$ and none of the other symbols is 0. Then, there are $p$ distinct $p$-words with 0 appearing in precisely one of the $p$ slots. Thus, $C_p = p$. Similarly, for $n$-words with $n > p$, the first 0 must appear in any one of the first $p$ slots. Once this slot is chosen, the rest of the word is uniquely determined by periodicity. Thus, we have $p$ distinct $n$-words, so that $C_n = p$ for $n \ge p$. This result also applies to UHE periodic orbits that do not pass through the origin.

Since it vanishes, topological entropy does not distinguish between periodic trajectories of different periods. However, the plateauing of word-complexity at the period $p$ does encode some of this information.

\vspace{5pt}

\fl \textbf{Word-frequency for periodic sequences:} Recall from \S \ref{s:seq-space-shift-map-word-stat} that the word frequency $f_w$ of a word is the frequency with which it occurs in a symbol sequence. Suppose $s$ is a $p$-periodic sequence starting with the symbol 0. For example, the UHE orbit with slope $3/2$ $s(\g_{3/2}) = 02123212 \, 02123212 \cdots$ has period eight. For 1-letter words, $f_0 = f_3 = 1/8$, $f_1 = 1/4$ and $f_2 = 1/2$. Similarly, for 2-words, $f_{02} = f_{23} = f_{32} = f_{20} = 1/8$, $f_{21} = f_{12} = 1/4$. Notice that the sum of frequencies of $k$-words is unity: $\sum_{|w|=k} f_w(s) = 1$. For $k < p$, the word-frequencies of some $k$-words may exceed $1/p$ due to repetitions since the word-complexity $C_k$ may be less than $p$. On the other hand, as noted above, for $k \ge p$, there are $p$ distinct $k$-words ($C_{k \ge p} = p$) labeled by the position of the first 0 in the words. Therefore, $f_w(s) = 1/p$ for all words of length $|w| \ge p$ that appear in the sequence.

%-----------------
\subsection{Quasiperiodic trajectories: Grammar and word statistics}
\label{s:uhe-quasiperiodic}
%-----------------

Next, we consider UHE quasiperiodic trajectories starting at the origin with irrational slope. Rotor exchange symmetries of \S \ref{s:structure-traj-sym-seq-UHE} allow us to restrict to $\sig > 1$.

%-----------------
\subsubsection{Word-complexity, branching and right-special words}
\label{s:conjectures-uhe-quasiperiodic}
%-----------------

Based on numerical experiments, we formulate some conjectures on the symbol sequences and word-complexity of UHE trajectories with irrational slope $\sig$. These are explained for asymptotic values of $\sig$ in \S \ref{s:asymp-argue-Cn-formula} and for general $\sig$ via a connection to Sturmian sequences in \S \ref{s:sturmian}.

\vspace{5pt}

\begin{table}[b]
	\begin{center}
		\begin{tabular}{|c|c|c|c|c|c|c|c|c|c|}
			\hline
			\boldmath$n$ & \multicolumn{9}{|c|}{\textbf{\boldmath $n$-letter words in symbol sequence of UHE orbit with slope $\sig = \sqrt{2}$}}\\
			\hline
			1 & 0 & \multicolumn{2}{|c|}{1} & \multicolumn{4}{|c|}{2} & \multicolumn{2}{|c|}{3} \\
			\hline 2 & 02 & \multicolumn{2}{|c|}{12} & \multicolumn{2}{|c|}{21} & \multicolumn{2}{|c|}{23} & \multicolumn{2}{|c|}{32} \\
			\hline
			3 & 021 & 121 & 123 & \multicolumn{2}{|c|}{212} & \multicolumn{2}{|c|}{232} & \multicolumn{2}{|c|}{321} \\
			\hline
			4 & 0212 & 1212 & 1232 & 2121 & 2123 & \multicolumn{2}{|c|}{2321} & \multicolumn{2}{|c|}{3212} \\
			\hline
			5 & 02123 & 12123 & 12321 & 21212 & 21232 & \multicolumn{2}{|c|}{23212} & 32121 & 32123 \\
			\hline
			6 & 021232 & 121232 & 123212 & 212123 & 212321 & 232121 & 232123 & 321212 & 321232\\
			\hline
		\end{tabular}
	\end{center}
	\caption{\small For $n = 1, \dots, 6$, the $n$th row contains a list of all distinct $n$-words in $s(\g_{\sqrt{2}})$. Each successive row has one more word than the previous one, resulting from two possible ways of extending one of the words in the previous row (branching or right-special word). The presence of only one such branching leads to the formula $C_n = n+3$. In fact, we have argued that the unique $n$-letter word starting with 0 read backwards (excluding the 0) coincides with the $(n-1)$-letter word that branches. For example, the word 0212 tells us that 212 is right-special and must branch. Similarly, the word 02123 tells us that 3212 is also right-special.}
	\label{t:Cn-sqrt2}
\end{table}

\begin{figure}[t]
	\centering
	\begin{subfigure}[b]{0.40\textwidth}
		\centering
		\includegraphics[width=\textwidth]{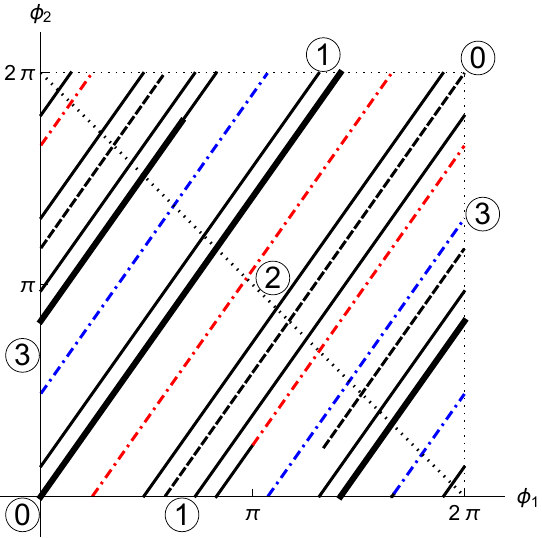}
		\caption{}
		\label{f:highE-sig-sqrt2}
	\end{subfigure}
	\hspace{0.025\textwidth}
	\begin{subfigure}[b]{0.40\textwidth}
		\centering
		\includegraphics[width=\textwidth]{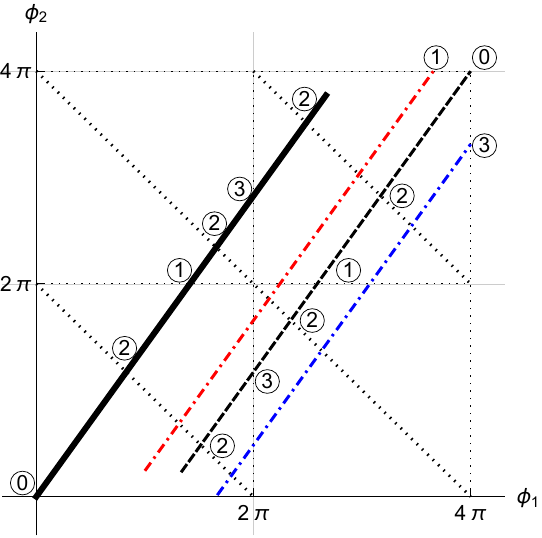}
		\caption{}
		\label{f:right-special-5word-sqrt-2}
	\end{subfigure}
	\caption{\small Quasiperiodic UHE trajectory passing through the origin at $t = 0$ with $\sig = \sqrt{2}$ for $\dot \vf_1 = 1$ and $\dot \vf_2 = \sqrt{2}$. (a) About twelve $\vf_2$ cycles are shown on the $2\pi \times 2\pi$ $\vf_1$-$\vf_2$ fundamental domain for $t \in [-8\pi/3, 10\sqrt{2}\pi]$. Segments for $[0,8 \pi/3]$ (thick black), for $[-8\pi/3, 0]$ (dashed black) and those that come close to it on the left (dot-dashed red $[3\pi,  4\sqrt{2} \pi]$) and on the right (dot-dashed blue $[17\pi/3, 8\pi]$) are highlighted. Thin black lines correspond to the trajectory at intermediate times. (b) The same highlighted segments of $\g_{\sqrt{2}}$ are shown on a $4\pi \times 4 \pi$ domain. We see that the right-special 5-word 23212 corresponding to the dashed black line branches into two 6-words 232121 and 232123 along the red and blue dot-dashed lines.}
	\label{f:sig-sqrt-2}
\end{figure}

\fl \textbf{Conjecture on word-complexity.} Using (\ref{e:uhe-linear-evol-rel-angles}) and the rules of \S \ref{s:UHE-ss}, we generate symbol sequences of length $N$ (in the 0123 alphabet) for UHE trajectories $\g_\sig$ starting at the origin with irrational slope $\sig > 1$. For example, when $\sig = \sqrt 2$ (see Fig.~\ref{f:highE-sig-sqrt2}), the symbol sequence is \small
	\beq
	s(\g_{\sqrt{2}}) = 02 \, 12321232 \, 12 \, 12321232 \, 12 \, 123212321232 \, 12 \, 12321232 \, 12 \, 123212321232 \, 12 \, 12321232 \cdots
	\label{e:seq-gamma-sqrt-2}
	\eeq \normalsize
We find all $n$-letter words in $s(\g_\sig)$ and count $C_n$ for $n \le N$ (see Table \ref{t:Cn-sqrt2}). For all irrational numbers we have checked, we find that $C_n = n+3$ until finite-$N$ truncation errors set in ($C_n = n+3$ holds up to larger $n$ when $N$ is increased). Thus, we conjecture that $C_n = n+3$ for such UHE quasiperiodic trajectories starting at the origin. This proposed word-complexity is sublinear ($C_n \leq C n$ for all $n$, say, with $C = 4$) and subadditive ($C_{m+n} \le C_m + C_n$) and implies that UHE sequences have zero topological entropy and power entropy equal to one.

Recall from Table~\ref{t:rotor-exchanges} that the symbol sequence for a trajectory with $\sig < 1$ can be obtained from that of an appropriate $\sig > 1$ trajectory by exploiting rotor exchanges. For instance, the sequences for $\sig' = 1/\sqrt{2}$ and $\sig'' = -(1 + \sqrt{2})$ are got from that for $\sig = \sqrt{2}$ by swapping $1 \leftrightarrow 3$ and $1 \leftrightarrow 2$ respectively. Thus, the proposed formula $C_n = n+3$ applies to all irrational $\sig$. 

\iffalse
In particular, the 3-words for these values of $\sig$ are:
\beqs
\sig = \sqrt{2} :& 021, 121, 123, 212, 232, 321 \cr
\sig = 1/\sqrt{2} :& 023, 323, 321, 232, 212, 123 \cr
\sig = -(1 + \sqrt{2}) :& 012, 212, 213, 121, 131, 312.
\eeqs
4-letter words 
$\sig = \sqrt{2}$ 0212, 1212, 1232, 2121, 2123, 2321, 3212
$\sig = -(1 + \sqrt{2})$ 0121, 2121, 2131, 1212, 1213, 1312, 3121
$\sig = 1/\sqrt{2}$ 0232, 3232, 3212, 2323, 2321, 2123, 1232 
\fi

\vspace{5pt}

\fl \textbf{Conjecture on branching and right-special words.} The possible $n$-words in a symbol sequence $s(\g_\sig)$ are obtained by suffixing each $(n-1)$-word by one letter. Given that there are four $1$-words ($C_1 = 4$), the formula $C_n = n+3$ can be explained if precisely one of the $(n-1)$-words admits two distinct one-letter suffixes (1 and 3 for $\sig > 1$), while the others are extendable in only one way (see Table~\ref{t:Cn-sqrt2}). Now, a subword $w$ of $s(\g)$ is called right-special if there are at least two symbols $a, b$ such that both $wa$ and $wb$ are subwords of $s(\g)$ (\S 1.1 of Ref.~\cite{fogg-book-2002}). Then our conjecture that only one $n$-word in $s(\g_\sig)$ branches in two ways, while all other $n$-words are extendable uniquely to get $(n+1)$-words, implies there is a unique right-special $n$-word in $s(\g_\sig)$ for each $n = 1, 2, \dots$ and $\sig$ irrational. In other words, $C_{n+1} - C_n$, which is equal to the number of right-special $n$-words (each of which branches in precisely two ways), must equal unity.

\vspace{5pt}

\fl \textbf{Identifying the right-special word.} Based on a study of the examples $\sig = \sqrt{2}$ and $\sqrt{3}$, we propose that the $(n-1)$-word that branches (in two ways) coincides with the unique $n$-word starting with 0 read backwards (excluding the 0). For instance, when $\sig = \sqrt{2}$, the 6-letter word 021232 tells us that the 5-letter word 23212 must branch into 232121 and 232123 (see Eq. (\ref{e:seq-gamma-sqrt-2}) and Table~\ref{t:Cn-sqrt2}). To see why this is true, we observe that a quasiperiodic trajectory passing through the origin at $t = 0$ must come arbitrarily close (on either side) to the segment that precedes the origin. In more detail, notice that 232120 (021232 read backwards) is the word corresponding to the segment of $\g_{\sqrt{2}}$ that occurs just before $t = 0$. Although this quasiperiodic trajectory will not return to the origin, it must come arbitrarily close to this segment (on either side) if we wait long enough. Thus, $\g_{\sqrt{2}}$ must include segments that hit the ceiling or wall instead of passing through the origin (see Fig.~\ref{f:right-special-5word-sqrt-2}). These segments will correspond to the words 232121 and 232123. Evidently, this argument applies to any irrational $\sigma$. What is more, branching can only occur in a word associated with a segment of the trajectory that ends at the origin. This is because the origin is the only coincidence point that has both a wall and a floor arbitrarily close to it. This explains the uniqueness of the right-special $n$-word for each $n$.

\vspace{4pt}

\fl \textbf{Approximating UHE orbits by high-energy orbits.} Suppose we try to numerically approximate a UHE straight line orbit $\g_\sig$ with a high but finite energy $E$ orbit with the same initial location and initial slope $\dot \vf_2 (0) / \dot \vf_1 (0) = \sig$. Even if $\sig$ is rational, the resulting trajectory is not periodic, but behaves like a UHE quasiperiodic orbit; its symbol sequence has word-complexity $n+3$. Thus, it is difficult to numerically approximate a UHE periodic orbit in this manner. On the other hand, if $\sig$ is irrational, the symbol sequence of the resulting high energy trajectory matches an initial portion of that of the corresponding UHE orbit. What is more, by increasing $E$, the fluctuations in $\dot \vf_1$ and $\dot \vf_2$ decrease and the matching portion of the sequence can be made arbitrarily long. Furthermore, the word-complexity of the approximate sequence is $n+3$ and matches that of the UHE sequence. Similarly, the word-frequencies $f_w$ (see \S \ref{s:word-freq-uhe-quasiper}) of the approximate symbol sequence approach those of the UHE sequence. Thus, we can numerically approximate UHE quasiperiodic orbits and their symbol sequences with finite energy ones. However, the results and conjectures reported in \S \ref{s:uhe-periodic} and \S \ref{s:uhe-quasiperiodic} are not based on numerical solutions at high energy but obtained by finding the rotor-coincidence times from the analytic solution (\ref{e:uhe-linear-evol-rel-angles}).

%-----------------
\subsubsection{Grammar rules for ultra-high-energy symbol sequences}
\label{s:gram-rule-uhe}
%-----------------

Here, we attempt to formulate `grammar rules' for constructing words (in the 0123 alphabet) that can appear in the symbol sequences $s(\g_\sig)$ of UHE trajectories starting at the origin with slope $\sig > 1$. Among other things, these rules will allow us to explain the formula $C_n = n+3$ for large and small irrational $\sig$ in \S \ref{s:asymp-argue-Cn-formula}.

\fl \textbf{(1)} The words should contain alternating 2s. This is clear from Fig.~\ref{f:highE-sig-3/2}: 2s arise from intersections with the 1-3 coincidence diagonal.

% For this reason, one could omit all 2s and record only the three symbols 0, 1 and 3. However, we do not do this here since this rule only applies to the UHE trajectories with $\sig>1$ and doing so would come in the way of comparison with symbol sequences in other regimes.

\fl \textbf{(2)} Two consecutive letters in a word cannot be identical. 

% This is equivalent to the statement that the 2-words 00, 11, 22 and 33 are forbidden.

\vspace{5pt}

\begin{figure}[ht]
	\centering
	\begin{subfigure}[b]{0.3\textwidth}
	\centering
	\includegraphics[width=0.75\textwidth]{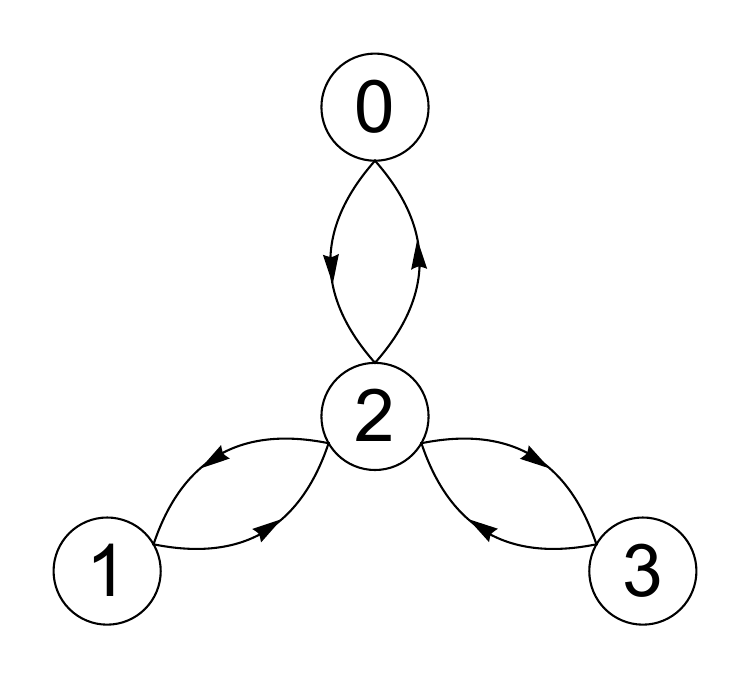}
	\caption{}
	\label{f:mg-v-12}
	\end{subfigure}
	\begin{subfigure}[b]{0.3\textwidth}
	\centering
	\includegraphics[width=0.9\textwidth]{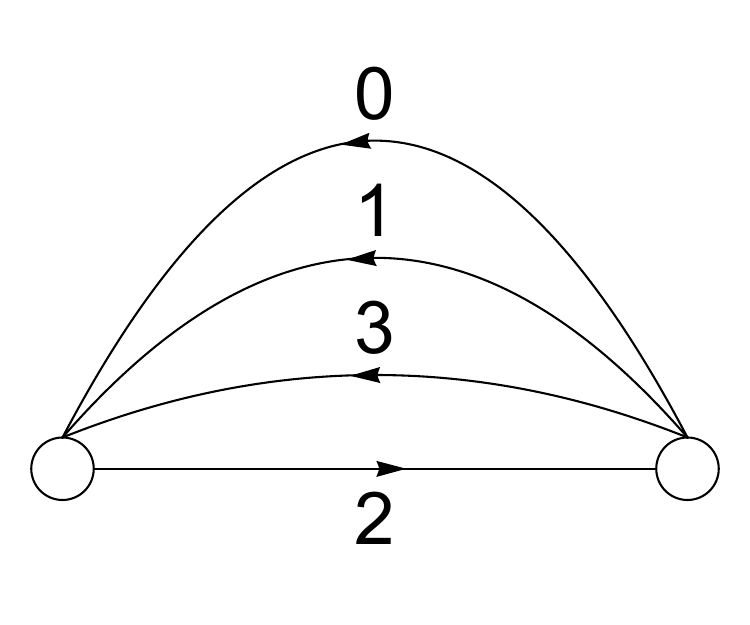}
	\caption{}
	\label{f:mg-e-12}
	\end{subfigure}
	\begin{subfigure}[b]{0.36\textwidth}
	\centering
	\includegraphics[width=0.9\textwidth]{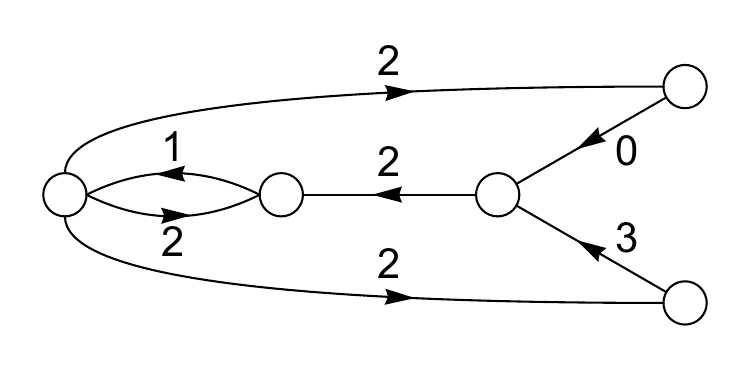}
	\caption{}
	\label{f:mg-e-123}
	\end{subfigure}
\caption{\small(a) Vertex-labeled Markov graph for grammar rules \textbf{(1)} and \textbf{(2)} has four vertices $v_0, v_1, v_2, v_3$ corresponding to the symbols 0, 1, 2, 3 with bidirectional edges connecting $v_2$ to every other vertex. (b, c) Edge-labeled Markov graphs for rules \textbf{(1,2)} and \textbf{(1,2,3)}. To incorporate rule \textbf{(4)} for irrational $\sig$, we remove the top edge in (c) with label 2.}
	\label{f:mg-123}
\end{figure}

\fl \textbf{Adjacency matrix and Markov graph.} Let $s_j$ be any letter in the alphabet ${\cal A} = \{0, 1, 2, 3\}$. If \textbf{(1), (2)} were the only rules and all sequences $s = \cdots s_j \cdots$ satisfying them could arise, then the space of such symbol sequences would be determined by the 0-1 transition (or adjacency) matrix $A = \begin{smmat}
	0 & 0 & 1 & 0 \cr
	0 & 0 & 1 & 0 \cr
	1 & 1 & 0 & 1 \cr
	0 & 0 & 1 & 0 
	\end{smmat}$,
with rows and columns labeled by the symbols 0, 1, 2 and 3. This means the only 2-words $s_j s_{j+1}$ that can appear in a symbol sequence are those for which the adjacency matrix elements $A_{s_j s_{j+1}} = 1$ (\S~1.9 of Ref.~\cite{katok-hasselblatt-1995}). The associated vertex- and edge-labeled Markov graphs  are shown in Figs.~\ref{f:mg-v-12} and \ref{f:mg-e-12}. If these were the only rules, the word-complexity $C_{n+1}$ would be the sum of the entries of $A^n$. However, there are additional grammar rules.

\fl \textbf{(3)} For $\sig > 1$, a trajectory starting on the wall must hit the ceiling before encountering the wall again, as in Fig.~\ref{f:highE-sig-3/2}. Consequently, 02 and 32 must be followed by 1. Rules \textbf{(1)}, \textbf{(2)} and \textbf{(3)} may together be represented by the edge-labeled graph of Fig.~\ref{f:mg-e-123}.

\fl \textbf{(4)} Here, we consider symbol sequences that begin with a 0. (a) For $\sig$ irrational, 0 appears only at the beginning, so there is only one word of any given length containing 0. The edge-labeled Markov graph incorporating rule \textbf{(4)} may be obtained from Fig.~\ref{f:mg-e-123} by removing the top edge carrying the symbol 2. (b) For rational $\sig$, the block between successive 0s is the repeating word. Therefore, if a word ends with 0, there is only one way of extending it to lengthier words, i.e., there can be no branching.

The next two rules concern the possible number of repetitions of 12 or 1232 arising from consecutive unbroken and broken lines introduced in \S \ref{s:structure-traj-sym-seq-UHE} and Fig.~\ref{f:highE-sig-3/2}.

\vspace{4pt}

\fl \textbf{(5)} Recall that an unbroken line corresponds to the words 021, 120 or 121. For irrational $\sig > 1$, the number of consecutive unbroken lines is either $\lfloor \sig \rfloor - 1$ or $\lfloor \sig \rfloor$. This constrains the number of repetitions of 12. For instance, the sequence must start with $02$ followed by $\lfloor \sig \rfloor$ repetitions of 12 suffixed by 32. Then onwards, there can only be either $\lfloor \sig \rfloor$ or $\lfloor \sig \rfloor + 1$ repetitions of 12 prefixed and suffixed by 32. Nonintegral rational $\sig > 1$ allows for one more possibility: $\lfloor \sig \rfloor$ repetitions of 12 on either side of 02, prefixed and suffixed by $32$. This corresponds to $2 \lfloor \sig \rfloor$ consecutive unbroken lines as illustrated by the symbol sequences for 
	\beqs
	\sig = 3/2: && \cdots 02 1232 \boldsymbol{12 02 12}32 12 0 \cdots \quad \text{and for} \cr
	\sig = 8/5: && \cdots 02 1232 \boldsymbol{12 12}32 1232 12 1232 \boldsymbol{12 02 12}32 12 1232 1232 12 1232 12 0 \cdots.
	\eeqs
Finally, for integral $\sig > 1$, the repeating block is 02 followed by $\sig - 1$ repetitions of 12. In particular, the symbol 3 is absent as there are no wall crossings. 

\vspace{4pt}

\fl \textbf{(6)} Recall that a broken full line corresponds to the word 12321. For integer $\sig > 1$, there are no broken lines, so the block $12321$ is forbidden. For nonintegral $\sig > 2$, broken lines are isolated, so the word 12321 is always prefixed by 12 and suffixed by 21. For $1 < \sig < 2$, the number $N_b$ of consecutive broken lines can take at most two values: $M - 1 \le N_b \le M$ where $M = \lfloor \sig / (\sig - 1) \rfloor_s$. Thus, repetitions of 1232 must come in groups of either $M-1$ or $M$, flanked by 02 or 12 on the left and by 120 or 121 on the right. For irrational $\sig$, these bounds are achieved if we wait long enough. For example, if $\sig = \sqrt{2}$, $N_b$ can be either 2 or 3, both of which are achieved (see Fig.~\ref{f:sig-sqrt-2}). For rational $\sig = p/q > 1$ (with $p,q$ coprime), the upper bound $M$ is achieved only if $p < (M q - 1)/(M - 1)$.

\vspace{5pt}

\fl \textbf{Need for more rules: Example of $\boldsymbol{\sig = \sqrt{2}}$.} For $\sig = \sqrt{2}$, rules \textbf{(1)-(6)} reproduce the formula $C_n = n+3$ for $n \le 22$ but overestimate $C_{23}$ by one word. As noted in \S \ref{s:conjectures-uhe-quasiperiodic}, for $C_{n+1} = C_n + 1$ to hold, only one $n$-word can branch (be extended by suffixing 1 or 3). The rules \textbf{(1)-(6)}, however, allow for two branchings at $n=22$. This implies there are additional grammar rules that enforce the uniqueness of right-special $n$-words.

\vspace{5pt}

\fl \textbf{Not a topological Markov chain.} We consider the space $X_+$ with left shift $\tau$ corresponding to all forward time translates of the sequence of a UHE orbit that starts at the origin with fixed irrational slope $\sig > 1$. A subshift is said to be of finite type or a topological Markov chain if it may be defined via an adjacency matrix or equivalently via an $(n+1)$th rank adjacency tensor (called an $n$-step Markov chain, see \S~1.9 of Ref.~\cite{katok-hasselblatt-1995}). A subshift of finite type (SFT) may also be viewed as consisting of all sequences that do not contain a finite number of forbidden words \cite{bruin-book-2022}. If \textbf{(1)} and \textbf{(2)} were the only rules, $(X_+,\tau)$ would be a topological Markov chain. In fact, rules \textbf{(1)} and \textbf{(2)} are equivalent to the statement that the words 00, 01, 03, 10, 11, 13, 22, 30, 31 and 33 are forbidden. Furthermore, rule \textbf{(3)} disallows the 3-words 020, 023, 320 and 323, although they are allowed by \textbf{(1)} and \textbf{(2)}. It is noteworthy that rules \textbf{(1), (2)} and \textbf{(3)} can together be formulated as a 2-step topological Markov chain, corresponding to a third-rank 0-1 adjacency tensor with $A_{ijk} = 0$ for all forbidden 3-words $ijk$. It should be possible to formulate the grammar rules including \textbf{(5)} and \textbf{(6)} in terms of an adjacency tensor whose rank depends on $\sig$. 

However, the conjectured absence of more than one branching suggests there must be grammar rules involving arbitrarily long words or having arbitrarily long `memory'. This indicates that $(X_+,\tau)$ is not an SFT. This is consistent with the connection to Sturmian sequences which will be established in \S \ref{s:sturmian}, since it is known that Sturmian shifts are not subshifts of finite type \cite{lind-marcus-book-1995,fogg-book-2002,bruin-book-2022,lothaire-2002}. Note that for $\sig = p/q$ rational, the subshift corresponding to the UHE periodic sequence $s_\sig$ with repeating block $b$ is an SFT. In fact, it may be defined in terms of a $0$-$1$ adjacency tensor $A$ of rank $N_{p/q}$ (\ref{e:num-lett-1-period}), with $A_w = 1$ only if $w$ is a cyclic permutation of $b$.

%-----------------
\subsubsection[Asymptotic arguments for word-complexity formula $C_n = n+3$]{Asymptotic arguments for word-complexity formula \texorpdfstring{$\boldsymbol{C_n = n+3}$}{Cn=n+3}}
\label{s:asymp-argue-Cn-formula}
%-----------------

Here, we provide arguments for the conjectured UHE word-complexity formula $C_n = n + 3$ of \S \ref{s:conjectures-uhe-quasiperiodic} in the asymptotic regimes $\sig \gg 1$ and $\sig \gtrsim 1$.

\vspace{5pt}

\fl \textbf{Asymptotically large slope.} For $\sig \gg 1$, trajectories are nearly vertical and hit the floor/ceiling many times before encountering a wall, resulting in sequences with far-separated 3s. We will assume that $n$ is not too large ($n \lesssim 2 \lfloor \sig \rfloor$ is adequate by rule \textbf{(5)} of \S \ref{s:gram-rule-uhe}), so that any $n$-letter word can contain at most one 3. Consider, for example, $n = 5$. For sufficiently large $\sig$, the unique 5-letter word starting at the origin is 02121. In addition, there are two more 5-letter words without a 3: 12121 and 21212, which correspond to sections of the trajectory that lie strictly between the walls. Next, we identify words that contain a 3. Note that the word starting with 0 cannot contain a 3 because of our assumptions on $\sig$ and $n$. Thus, 5-letter words that contain a 3 are obtained by replacing any one of the 1s in the latter two words: 12121 and 21212. There are five possible ways of doing this, resulting in the words 32121, 12321, 12123, 23212 and 21232. Thus, we get a total of $5+3 = 8$ five-letter words, in agreement with the proposed formula $C_n = n+3$. 

More generally, there are three $n$-words that do not contain a 3: $0212 \cdots$, $1212 \cdots$ and $2121 \cdots$. The $n$-letter words containing one 3 are obtained by replacing any one of the 1s in the latter two words. Since there are $n$ such 1s that can be replaced, we get $n$ $n$-words containing a 3. Thus, $C_n = n+3$ under these circumstances. It should be possible to extend this argument to larger values of $n$ by bearing in mind that the $n$-words may have to contain one or more 3s due to encounters with walls. In effect, more 1s will be replaced by 3s although the total number of words will still be $n+3$.  

\vspace{5pt}

\fl \textbf{Slope close to one.} Suppose $\sig = 1 + \eps$ for $0 < \eps < 1$. Then, we cannot have two consecutive unbroken lines (each with sequence 021 or 121) since the maximum number of such lines is $\lfloor \sig \rfloor$ from rule \textbf{(5)} of \S~\ref{s:gram-rule-uhe}. On the other hand, by making $\eps$ sufficiently small, we can have arbitrarily many consecutive broken lines. In fact, from rule \textbf{(6)}, the number of consecutive broken lines is either $\lfloor 1/\eps \rfloor_s$ or $\lfloor 1 + 1 / \eps \rfloor_s$. For instance, three consecutive broken lines starting on the floor and ending on the ceiling correspond to the string 1 232 1 232 1 232 1. Thus, the symbol sequences consist of isolated unbroken lines separating several consecutive broken lines. In particular, for $0 < \eps < 1/2$, the sequence must start with an unbroken line followed by two or more broken lines: 0 2 1 232 1 232 1 $\cdots$. 

Let us fix $\eps$ and assume that $n$ is not too large so that an $n$-word cannot overlap two unbroken lines (i.e., $n$ is  smaller than the length of the sequence corresponding to the minimum number of consecutive broken lines). This places the upper bound $n < 4 \times \lfloor 1/\eps \rfloor_s$. In this case, for $n \geq 2$ the $n$-words are of three sorts: (i) the unique word that starts with a zero, 0212321232$\cdots$, (ii) four words that do not have a zero and do not contain the block 121 corresponding to an unbroken line (i.e., only involve broken lines): 12321232$\cdots$, 23212321$\cdots$, 32123212$\cdots$ and 21232123$\cdots$, and (iii) $n-2$ words that do not have a zero but contain 121 (since the block 121 can be placed in $n-2$ ways in an $n$-word). Hence, there are $1 + 4 + (n-2) = n+3$ distinct $n$-letter words, for $n \geq 2$. Moreover, it is clear that there are four 1-letter words, so $C_1 = 4$. Thus, we get the conjectured word-complexity in this limit. For example, the distinct 5-letter words of the above three sorts are (i) 02123, (ii) 12321, 23212, 32123, 21232 and (iii) 32\textbf{121}, 2\textbf{121}2, \textbf{121}23.

%----------------
\subsubsection{Connection to Sturmian sequences}
\label{s:sturmian}
%----------------

Here, we present a correspondence between UHE quasiperiodic rotor-coincidence sequences and Sturmian sequences and use properties of the latter to deduce that the word-complexity of the former is indeed given by $C_n = n+3$, as conjectured in \S \ref{s:conjectures-uhe-quasiperiodic}.

\vspace{5pt}

\fl {\bf Sturmian sequences} (Chapter 6 of Ref.~\cite{fogg-book-2002}, Section 4.3 of Ref.~\cite{bruin-book-2022} and Chapter 2 of Ref.~\cite{lothaire-2002}) are those for which the word-complexity $C^s_n = n+1$ for each $n \ge 0$. They are the aperiodic binary sequences (i.e., over a two-letter alphabet) with minimal word-complexity \cite{morse-hedlund-1938,morse-hedlund-1940}. We will relate our UHE quasiperiodic sequences to Sturmian sequences. Let us begin by recalling that Sturmian sequences may be generated as ``cutting sequences'' on a square grid. Consider the straight line free particle trajectory $y = \al x + \rho$ drawn on the $x$--$y$ plane with grid lines marked at integer values of $x$ and $y$. We assume that $\al$ is irrational. If $\rho = a\alpha + b$ for some integers $a$ and $b$, then the trajectory passes through precisely one grid point, namely $(x = -a,\, y = b)$. If $\rho$ is not of this form, the trajectory does not pass through any grid point. Of relevance to us is the case $\rho = 0$ so that the trajectory may be regarded as starting at the origin $(0, 0)$. Leaving aside the initial point, we record symbols 1 and 3 when the trajectory subsequently cuts the horizontal and vertical grid lines. The resulting symbol sequence is Sturmian with irrational slope $\al$ and intercept 0. More generally, one gets a Sturmian sequence with slope $\al$ and intercept $\rho$. It is known that Sturmian sequences do not form a subshift of finite type  (see Theorem 3.14 of Ref.~\cite{bruin-book-2022} and Refs.~\cite{lind-marcus-book-1995,lothaire-2002}).

\vspace{5pt}

\fl {\bf Mapping to UHE rotor-coincidence sequences.} To make the connection to our UHE trajectories starting at the origin with irrational slope $\sig > 1$, we choose the relative angles as $\vf_1 = 2 \pi x$, $\vf_2 = 2 \pi y$ and slope $\sig = \al$. Then, there is a one-to-one correspondence between our rotor-coincidence symbol sequences and Sturmian sequences. Given a Sturmian sequence $s_1 s_2 s_3 \cdots$, the corresponding rotor sequence is $02 s_1 2 s_2 2 s_3 2 \cdots$. We may use this correspondence to relate the word-complexity $C_n$ of UHE quasiperiodic rotor sequences to the word-complexity $C^s_n = n + 1$ of Sturmian sequences. We will do this by constructing rotor $n$-words from Sturmian words. 

To begin with, there are four rotor 1-words: 1, 3, 2 and 0, of which the first two are Sturmian while the last two are not. Thus, $C_1 = C^s_1 + 2 = 4$. Next, there are five rotor 2-words, which for $\sig > 1$ are 21, 23, 12, 32 and 02. The first four are obtained by prefixing or suffixing 2 to the two Sturmian 1-words 1 and 3, while the last word is the unique rotor 2-word starting with 0. Thus, $C_2 = 2 C^s_1 + 1 = 5$. Next, there are six rotor 3-words, which for $\sig > 1$ are 212, 232, 121, 123, 321 and 021. Aside from the unique word starting with 0, we notice that the first two words are obtained by prefixing and suffixing the symbol 2 to Sturmian 1-words, while the remaining three are got by inserting a 2 in the middle of each Sturmian 2-word. Thus, $C_3 = C^s_1 + C^s_2 + 1 = 6$. For $n \ge 5$, although the rotor $n$-words depend on the value of $\sig > 1$, they can, with one exception, be generated from Sturmian words of length roughly $n/2$. More precisely, for odd $n$, the rotor $n$-words (aside from the one starting with 0) are obtained by inserting 2s in between the symbols of Sturmian $((n+1)/2)$-words as well as by inserting 2s before, after and in between the symbols of Sturmian $((n-1)/2)$-words: 
	\beq
	s_1 2 s_2 2 \cdots 2 s_{\frac{n+1}{2}} \quad \text{and} \quad 2 s_1 2 s_2 2 \cdots 2 s_{\frac{n-1}{2}} 2.
	\eeq
Therefore, the number of rotor $n$-words is 
	\beq
	C_n = C^s_{\frac{n+1}{2}} + C^s_{\frac{n-1}{2}} + 1 = n + 3, \; \text{for odd $n > 1$}. 
	\eeq
In a similar fashion, for even $n$, the rotor $n$-words can be obtained from the Sturmian $(n/2)$-words:
	\beq
	2 s_1 2 s_2 2 \cdots 2 s_{\frac{n}{2}} \quad \text{and} \quad s_1 2 s_2 2 \cdots 2 s_{\frac{n}{2}} 2,
	\eeq
so that
	\beq
	C_n = 2 C^s_{\frac{n}{2}} + 1 = n + 3, \; \text{for even $n$}.
	\eeq
Combining, we get the word-complexity $C_n = n + 3$ for UHE quasiperiodic sequences.

%-----------------
\subsubsection{Word-frequency of ultra-high-energy quasiperiodic sequences}
\label{s:word-freq-uhe-quasiper}
%----------------- 

We briefly describe some features of the word-frequency of rotor-coincidence sequences of UHE quasiperiodic orbits. Suppose we consider sequences of length $N \gg 1$ for a fixed slope $\sig$. We find the word-frequencies of all the $n+3$ words of length $n$. The sum of these frequencies approaches one as $N \to \infty$. By studying examples, we observe that the word-frequencies rapidly approach asymptotic values and that they take at most three distinct nonzero values (the unique word starting with 0 has limiting frequency $f_0 = 0$). For instance, when there are three distinct frequencies $0 < f^1 < f^2 < f^3$, there are $n_i$ words with frequency $f^i$ for $i = 1,2,3$ satisfying the relations $\sum_{i=1}^{3} n_i = n + 2$ and $\sum_{i=1}^{3} n_i f^i = 1$. Furthermore, when there are three distinct word-frequencies, they satisfy the additional relation $f_1 + f_2 = f_3$. The presence of at most three distinct frequencies as well as this sum rule are shared by Sturmian sequences (a consequence of the three-gap theorem, see Theorem 4.70 of Ref.~\cite{bruin-book-2022} and Theorem 2.2.37 of Ref.~\cite{lothaire-2002}).

%-----------------
\section{Chaotic trajectories: symbol sequences and word-complexity}
\label{s:word-cx-global-chaotic-band}
%----------------- 

\begin{figure}[ht]
	\centering
	\includegraphics[width=0.45\textwidth]{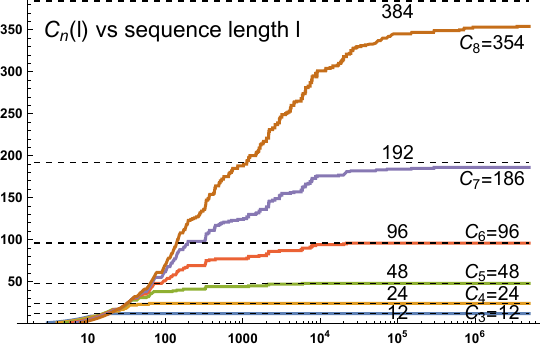}
	\caption{\small Word-complexity $C_n(l)$ plotted against the logarithm of the sequence length $l$ for words of length $n = 2,3, \cdots, 8$ in the symbol sequence of a chaotic trajectory with $E = 5.5$ and ICs $\{ \vf_1(0) = 0, \vf_2(0) = p_2(0) = 1 \}$ evolved up to $t = 6 \times 10^6$ in units where $m = r = g = 1$. The initial 0 in the symbol sequence is omitted. The horizontal lines are at the conjectured asymptotic word-complexities $C_n = 3 \times 2^{n-1}$. For $n \leq 6$, $C_n(l)$ is seen to saturate at this value while it is plausible that the curves for $n=7$ and $n=8$ are approaching this value.}
	\label{f:cn-vs-log-l}
\end{figure}

\fl \textbf{Sequences from the band of global chaos.} The trajectories we have considered so far are static, periodic or quasiperiodic and their symbol sequence word-complexities have been bounded or linearly growing, leading to vanishing topological entropy. We now briefly consider chaotic trajectories with a view to finding rotor-coincidence sequences with exponentially growing word-complexity and nonzero topological entropy. A natural place to look is in the energy band $5.33g \lesssim E \lesssim 5.6 g$, where there is numerical evidence for global chaos, ergodicity and mixing \cite{gsk-hs-2020,gsk-ay-2023}. To this end, we use the conditions of \S \ref{s:order-coinc-conditions} to extract the symbol sequences and associated word-complexities of such numerically generated chaotic trajectories. We find that two different numerical solutions with the same ICs in the globally chaotic energy band typically lead to trajectories with symbol sequences that disagree beyond a point (sequence length $\sim 100$ in our examples). Sensitivity to ICs magnifies the effect of round-off errors. Although the symbol sequences differ, we find that their word-complexities $C_n(l)$ for sequence length $l$ (see Fig.~\ref{f:cn-vs-log-l}) agree as long as $n \ll l$. On the other hand, round-off errors typically do not affect the symbol sequences of quasiperiodic trajectories (at least up to $l \sim 10^6$ in our examples).

\vspace{4pt}

\fl \textbf{Conjecture on word-complexity.} Aside from some exceptional ICs (e.g., $\vf_1(0) = \vf_2(0) = p_2(0) = 0$, corresponding to a periodic rotational breather), we find that the behavior of word-complexity is largely independent of ICs. In units where $m = r = g = 1$, Fig.~\ref{f:cn-vs-log-l} displays the word-complexity $C_n(l)$ for $n \leq 8$ as a function of sequence length $l$ for a trajectory with energy $E = 5.5$ evolved up to $t_{\rm max} = 6 \times 10^6$, corresponding to $l_{\rm max} \approx 6.72 \times 10^6$. For fixed word length $n$, we see that $C_n(l)$ increases with $l$ and appears to approach an asymptotic word-complexity. For $n = 1, 2, \ldots, 6$, this saturation seems to occur at the values $3,6,12,24,48,96$, achieved at sequence lengths $l = 5, 19, 52, 5891, 21775 \ll l_{\rm max}$. Remarkably, the asymptotic word-complexities for $n \leq 6$ match the formula $C_n = 3 \times 2^{n-1}$. Based on this, we conjecture that the word-complexity of coincidence symbol sequences of generic trajectories in the globally chaotic band is given by $C_n = 3 \times 2^{n-1}$, corresponding to a topological entropy $h = \log 2$ for these sequences. However, we see from Fig.~\ref{f:cn-vs-log-l} that the plateauing of $C_n(l)$ with increasing $l$ is rather slow so that $C_7(l)$ and $C_8(l)$ at $l = l_{\rm max}$ have only reached $186$ and $354$, which are lower than the conjectured values $192$ and $384$. We are inclined to attribute this to the effects of finite sequence length, which are more pronounced for larger $n$. 

\begin{figure}[ht]
	\centering
	\begin{subfigure}[b]{0.20\textwidth}
	\centering
	\includegraphics[width=0.7\textwidth]{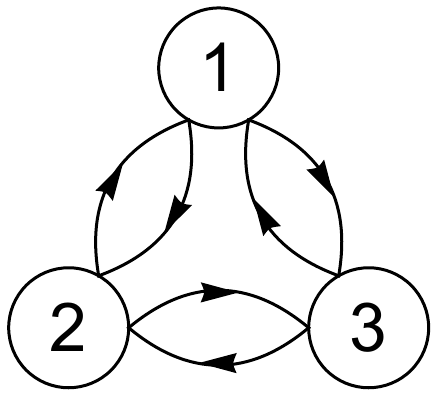}
%	\caption{}
	\label{f:markov-gr-vert-chaos}
	\end{subfigure}
	\begin{subfigure}[b]{0.20\textwidth}
	\centering
	\includegraphics[width=0.63\textwidth]{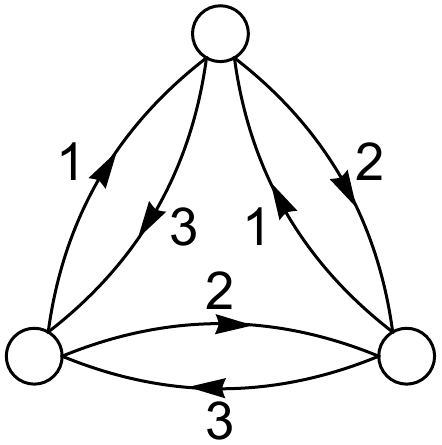}
%	\caption{}
	\label{f:markov-gr-edge-chaos}
	\end{subfigure}
\caption{\small Proposed vertex- and edge-labeled Markov graphs for chaotic sequences in the globally chaotic band.}
	\label{f:markov-graphs-global-chaos}
\end{figure}

\fl \textbf{Grammar rules and adjacency matrix.} Examination of the resulting symbol sequences of such chaotic trajectories leads us to the following observations. (i) Triple coincidences do not occur other than possibly at $t=0$ (which happens only if the IC involves a triple coincidence). In what follows, we only record symbols from coincidences at $t > 0$. This is because chaotic trajectories that begin at a triple coincidence and those that do not, appear to have the same word-complexity and frequencies if one omits the initial 0 symbol. The absence of triple coincidences in generic trajectories is to be expected since they require two vanishing conditions to be met, unlike pair coincidences (see \S \ref{s:order-coinc-conditions}). (ii) Repetitions of letters are not seen, i.e., the 2-letter words 11, 22 and 33 are forbidden while the others do occur. In the examples studied so far, repetition of symbols has only occurred in librational breathers below energy $4.5g$. We would expect repetition of symbols at higher energies to be either rare or absent due to a tendency to rotate rather than librate (see Figs.~\ref{f:A-C-regions-phi1-phi2-square} and \ref{f:highE-plot-1}). This should also apply in the band of global chaos which lies above this energy threshold. (iii) All 3-, 4-, 5- and 6-letter words without such repetitions do occur. 

Based on these observations, we infer that $C_1 = 3$, $C_2 = 2 C_1 = 6$, $C_3 = 2 C_2 = 12$, \dots, $C_6 = 2 C_5 = 96$. If there are no forbidden words that do not follow from (i) and (ii), then $C_{n+1} = 2 C_n$ since every $n$-word can be extended in precisely two ways while avoiding repetitions and the symbol 0. Elevating (i) and (ii) to grammar rules that hold in general would then imply the conjectured formula $C_n = 3 \times 2^{n-1}$. These rules may be encoded in the $4 \times 4$ adjacency matrix 
	\beq
	A = \begin{smmat} 0 & 0 & 0 & 0 \cr 0 & 0 & 1 & 1 \cr 0 & 1 & 0 & 1 \cr 0 & 1 & 1 & 0 \end{smmat},
	\eeq
with rows and columns labeled by the symbols 0, 1, 2 and 3. The corresponding vertex- and edge-labeled Markov graphs are shown in Fig.~\ref{f:markov-graphs-global-chaos}. Thus, if our conjecture holds, then generic coincidence symbol sequences in the globally chaotic regime may be modeled as a subshift of finite type (SFT) or topological Markov chain, unlike those at ultra-high energies. What is more, the conjectured absence of any grammar rules beyond the absence of zeros and repeated symbols reinforces the `fully chaotic' nature of this regime. Note that this adjacency matrix does not apply to the periodic orbits in this regime, which could have repeated triple coincidences.

\vspace{4pt}

\fl \textbf{Remark on the symbol 0.} Note that if the symbol 0 at the beginning of a chaotic trajectory that starts at a triple coincidence (but does not suffer one subsequently) is included, then the conjectured word-complexity increases by one: $C_n = 3 \times 2^{n-1} + 1$. Moreover, if we consider the corresponding bi-infinite sequence by extending backwards in time, then $C_n = 3 \times 2^{n-1} + n$. In other words, the inclusion of the symbol 0 from fine-tuned ICs has only a subleading effect on the word-complexity.

\vspace{4pt}

\fl \textbf{Word-frequency.} We also compute frequencies of words of length up to 4 occurring in symbol sequences of chaotic trajectories in this band. We find that all three 1-words occur with the same frequency 1/3, as do all six 2-words (1/6). Certain correlations emerge for longer words. Among the twelve 3-words, the six of the form $abc$ have a higher frequency ($\approx 0.095$) compared to the remaining six of the form $aba$ ($\approx 0.072$), where $a$, $b$ and $c$ stand for distinct symbols from the alphabet ${1, 2, 3}$. The twenty-four 4-words may be grouped into three frequency bins: the six $abab$ words with a frequency of $\approx 0.018$, the six $abca$ words with a frequency of $\approx 0.041$ and the twelve words of the form $abac$ or $abcb$ with a frequency of $\approx 0.054$. The forty-eight 5-words fall into six frequency groups: $ababa$ (0.0010), $abcab$ (0.0066), $ababc$ and $abcbc$ (0.017), $abaca$ (0.020), $abacb$ and $abcac$ (0.034), and $abcba$ (0.037). We hope to return to study the patterns in these frequencies elsewhere. 

\vspace{4pt}

\fl \textbf{Chaos outside the global band.} Just outside the globally chaotic band, say, at energy $E = 5$, the growth of word-complexity $C_n$ of coincidence sequences of chaotic trajectories is similar to that in the globally chaotic band (for $t_{\rm max} = 10000$ and $n$ up to 10). By contrast, for a chaotic trajectory well outside the band of global chaos (e.g., $E = 4.5$ with ICs $\vf_1(0) = \vf_2(0) = 0.5$ and $p_2(0) = 0$), the word-complexity seems to grow much slower than for chaotic trajectories in the globally chaotic band. 

%-----------------
\section{Discussion}
\label{s:discussion}
%----------------- 

In this paper, we have proposed a digitization of classical 3-rotor dynamics. By focusing on rotor coincidences, trajectories are replaced by symbol sequences in a four-letter alphabet. The word-complexity $C_n$ of such symbol sequences saturates at the period for periodic trajectories, is linear ($C_n = n + 3$) for ultra-high energy quasiperiodic orbits and is conjecturally given by $C_n = 3 \times 2^{n-1}$ for chaotic orbits in the band of global chaos. Some open questions have arisen from our work. (1) It would be interesting to understand the observed word frequencies of symbol sequences from the globally chaotic energy range based on our conjectured adjacency matrix. (2) Can we characterize all the periodic symbol sequences that occur in the globally chaotic regime and do they follow from the rule that simply forbids repetition of symbols? (3) What is the character of the word statistics for quasiperiodic orbits at low and intermediate energies? Preliminary investigations indicate interesting patterns in the word-complexity of such trajectories that are quite different from what we found at ultra-high energies. (4) What is the number of period-$n$ orbits in suitable families of symbol sequences, say, arising at ultra-high or low energies or in the globally chaotic band? For topological Markov chains, the Perron--Frobenius theorem provides information on this. (5) Is the stability of periodic trajectories encoded in the word statistics of the symbol sequences of nearby trajectories? (6) We note that the topological entropy we have found or conjectured for coincidence symbol sequences of periodic, UHE quasiperiodic and globally chaotic trajectories is to be distinguished from the topological  or Kolmogorov--Sinai entropy of the continuous 3-rotor dynamics. In particular, the rotor coincidence digitization does not have information on the time interval between successive coincidences.   Nevertheless, it would be interesting to examine to what extent entropies based on rotor coincidence codings can approximate these entropies of the Hamiltonian flow.

The sequences we have found at ultra-high energies and in the globally chaotic energy band are but two extreme cases for which we could find or conjecture relatively simple interpretations and explanations. We suspect that they are also the aperiodic 3-rotor coincidence sequences with minimal and maximal word-complexity. We expect to find richer symbolic dynamics in other regimes of the three-rotor problem and hope to return to the above questions elsewhere.

\vspace{5pt}

\fl {\bf Acknowledgements:} We would like to thank Himalaya Senapati for useful discussions and comments. 
We also thank the anonymous referees for their helpful suggestions. This work was supported in part by the Infosys Foundation and an INSPIRE-SHE scholarship of the Department of Science and Technology, Govt. of India.

%---------------

\end{document}